\documentclass[%
 reprint,
showpacs,
nofootinbib,
nobibnotes,
 amsmath,amssymb,
 aps,
prd,
hypertext,
]{revtex4-1}

\usepackage{graphicx}
\usepackage{dcolumn}
\usepackage{bm}
\usepackage{afterpage}

\usepackage{xspace}
\usepackage{color}
\usepackage{colortbl}
\usepackage{multirow}
\usepackage{amsmath}

\usepackage{ifthen} 
\newboolean{pdflatex}
\setboolean{pdflatex}{true} 
\newboolean{uprightparticles}
\setboolean{uprightparticles}{false} 
\usepackage{amssymb}
\usepackage{amsfonts}
\usepackage{upgreek}
\newboolean{articletitles}
\setboolean{articletitles}{true}

\def\lhcb {LHCb\xspace}
\def\ux85 {UX85\xspace}

\def\babar  {BABAR\xspace}
\def\belle  {BELLE\xspace}

\def\velo   {VELO\xspace}

\ifthenelse{\boolean{uprightparticles}}%
{

 \def\Ppi         {\ensuremath{\uppi}\xspace}

 \def\PDelta      {\ensuremath{\Delta}\xspace}                 
 \def\PXi      {\ensuremath{\Xi}\xspace}                 
 \def\PLambda      {\ensuremath{\Lambda}\xspace}                 
 \def\PSigma      {\ensuremath{\Sigma}\xspace}                 
 \def\POmega      {\ensuremath{\Omega}\xspace}                 
 \def\PUpsilon      {\ensuremath{\Upsilon}\xspace}                 
 
 \def\PB      {\ensuremath{\mathrm{B}}\xspace}                 
                  
 \def\PD      {\ensuremath{\mathrm{D}}\xspace}

 \def\PK      {\ensuremath{\mathrm{K}}\xspace}

 \def\Pb      {\ensuremath{\mathrm{b}}\xspace}                 
 \def\Pc      {\ensuremath{\mathrm{c}}\xspace}

 \def\Pi      {\ensuremath{\mathrm{i}}\xspace}

 \def\Pp      {\ensuremath{\mathrm{p}}\xspace}

 \def\Ps      {\ensuremath{\mathrm{s}}\xspace}

}
{

 \def\Ppi         {\ensuremath{\pi}\xspace}

 \mathchardef\PDelta="7101
 \mathchardef\PXi="7104
 \mathchardef\PLambda="7103
 \mathchardef\PSigma="7106
 \mathchardef\POmega="710A
 \mathchardef\PUpsilon="7107
                  
 \def\PB      {\ensuremath{B}\xspace}                 
                  
 \def\PD      {\ensuremath{D}\xspace}

 \def\PK      {\ensuremath{K}\xspace}

 \def\Pb      {\ensuremath{b}\xspace}                 
 \def\Pc      {\ensuremath{c}\xspace}

 \def\Pi      {\ensuremath{i}\xspace}

 \def\Pp      {\ensuremath{p}\xspace}

 \def\Ps      {\ensuremath{s}\xspace}

}

\def\squark    {\ensuremath{\Ps}\xspace}

\def\pion  {\ensuremath{\Ppi}\xspace}

\def\pip   {\ensuremath{\pion^+}\xspace}
\def\pim   {\ensuremath{\pion^-}\xspace}

\def\kaon  {\ensuremath{\PK}\xspace}
  \def\Kbar  {\kern 0.2em\overline{\kern -0.2em \PK}{}\xspace}

\def\Kz    {\ensuremath{\kaon^0}\xspace}
\def\Kzb   {\ensuremath{\Kbar^0}\xspace}
\def\KzKzb {\ensuremath{\Kz \kern -0.16em \Kzb}\xspace}
\def\Kp    {\ensuremath{\kaon^+}\xspace}
\def\Km    {\ensuremath{\kaon^-}\xspace}

\def\KpKm  {\ensuremath{\Kp \kern -0.16em \Km}\xspace}

  \def\Dbar    {\kern 0.2em\overline{\kern -0.2em \PD}{}\xspace}
\def\D       {\ensuremath{\PD}\xspace}

\def\Dz      {\ensuremath{\D^0}\xspace}
\def\Dzb     {\ensuremath{\Dbar^0}\xspace}
\def\DzDzb   {\ensuremath{\Dz {\kern -0.16em \Dzb}}\xspace}
\def\Dp      {\ensuremath{\D^+}\xspace}
\def\Dm      {\ensuremath{\D^-}\xspace}

\def\DpDm    {\ensuremath{\Dp {\kern -0.16em \Dm}}\xspace}

\def\Dstarp  {\ensuremath{\D^{*+}}\xspace}

\def\Dsp     {\ensuremath{\D^+_\squark}\xspace}

  \def\Bbar    {\kern 0.18em\overline{\kern -0.18em \PB}{}\xspace}

  \def\Y#1S{\ensuremath{\PUpsilon{(#1S)}}\xspace}

\def\proton      {\ensuremath{\Pp}\xspace}

\def\to                 {\ensuremath{\rightarrow}\xspace}

\def\CP                {\ensuremath{C\!P}\xspace}

\def\FL       {\ensuremath{F_{\mathrm{L}}}\xspace}
\def\AT#1     {\ensuremath{A_{\mathrm{T}}^{#1}}\xspace}           

\def\C#1      {\ensuremath{\mathcal{C}_{#1}}\xspace}                       
\def\Cp#1     {\ensuremath{\mathcal{C}_{#1}^{'}}\xspace}                    
\def\Ceff#1   {\ensuremath{\mathcal{C}_{#1}^{\mathrm{(eff)}}}\xspace}        
\def\Cpeff#1  {\ensuremath{\mathcal{C}_{#1}^{'\mathrm{(eff)}}}\xspace}       
\def\Ope#1    {\ensuremath{\mathcal{O}_{#1}}\xspace}                       
\def\Opep#1   {\ensuremath{\mathcal{O}_{#1}^{'}}\xspace}                    

\newcommand{\tev}{\ensuremath{\mathrm{\,Te\kern -0.1em V}}\xspace}
\newcommand{\gev}{\ensuremath{\mathrm{\,Ge\kern -0.1em V}}\xspace}
\newcommand{\mev}{\ensuremath{\mathrm{\,Me\kern -0.1em V}}\xspace}
\newcommand{\kev}{\ensuremath{\mathrm{\,ke\kern -0.1em V}}\xspace}
\newcommand{\ev}{\ensuremath{\mathrm{\,e\kern -0.1em V}}\xspace}
\newcommand{\gevc}{\ensuremath{{\mathrm{\,Ge\kern -0.1em V\!/}c}}\xspace}
\newcommand{\mevc}{\ensuremath{{\mathrm{\,Me\kern -0.1em V\!/}c}}\xspace}
\newcommand{\gevcc}{\ensuremath{{\mathrm{\,Ge\kern -0.1em V\!/}c^2}}\xspace}
\newcommand{\gevgevcccc}{\ensuremath{{\mathrm{\,Ge\kern -0.1em V^2\!/}c^4}}\xspace}
\newcommand{\mevcc}{\ensuremath{{\mathrm{\,Me\kern -0.1em V\!/}c^2}}\xspace}

\def\gsim{{~\raise.15em\hbox{$>$}\kern-.85em
          \lower.35em\hbox{$\sim$}~}\xspace}
\def\lsim{{~\raise.15em\hbox{$<$}\kern-.85em
          \lower.35em\hbox{$\sim$}~}\xspace}

\def\pt         {\mbox{$p_{\rm T}$}\xspace}
\def\et         {\mbox{$E_{\rm T}$}\xspace}

\def\dllkpi     {\ensuremath{\mathrm{DLL}_{\kaon\pion}}\xspace}

\def\tell1  {TELL1\xspace}
\def\ukl1   {UKL1\xspace}

\usepackage{mciteplus}

\begin{document}
\mciteErrorOnUnknownfalse
\preprint{APS/123-QED}

\title{Search for \CP violation in $D^{+} \to K^{-}K^{+}\pi^{+}$ decays}
\author{
\begin{center}
R.~Aaij$^{23}$, 
B.~Adeva$^{36}$, 
M.~Adinolfi$^{42}$, 
C.~Adrover$^{6}$, 
A.~Affolder$^{48}$, 
Z.~Ajaltouni$^{5}$, 
J.~Albrecht$^{37}$, 
F.~Alessio$^{37}$, 
M.~Alexander$^{47}$, 
G.~Alkhazov$^{29}$
% ...

{\it{\footnotesize
% ...
$ ^{5}$Clermont Universit\'{e}, Universit\'{e} Blaise Pascal, CNRS/IN2P3, LPC, Clermont-Ferrand, France\\
$ ^{6}$CPPM, Aix-Marseille Universit\'{e}, CNRS/IN2P3, Marseille, France\\
% ...
$ ^{23}$Nikhef National Institute for Subatomic Physics, Amsterdam, Netherlands\\
% ...
$ ^{29}$Petersburg Nuclear Physics Institute (PNPI), Gatchina, Russia\\
% ...
$ ^{36}$Universidad de Santiago de Compostela, Santiago de Compostela, Spain\\
$ ^{37}$European Organization for Nuclear Research (CERN), Geneva, Switzerland\\
% ...
$ ^{42}$H.H. Wills Physics Laboratory, University of Bristol, Bristol, United Kingdom\\
% ...
$ ^{47}$School of Physics and Astronomy, University of Glasgow, Glasgow, United Kingdom\\
$ ^{48}$Oliver Lodge Laboratory, University of Liverpool, Liverpool, United Kingdom\\
% ...
}}
\end{center}
}

\begin{abstract}
A model-independent search for direct \CP violation in the Cabibbo suppressed decay $D^+ \to K^- K^+\pi^+$ in a sample
of approximately 370,000 decays is carried out. The data were collected by
the LHCb experiment in 2010 and correspond to an integrated luminosity
of 35~pb$^{-1}$. The normalized Dalitz plot distributions for
$D^+$ and $D^-$ are compared using four different binning schemes that
are sensitive to different manifestations of \CP violation. No evidence for \CP asymmetry is found. 
\end{abstract}

\vspace*{-1cm}
\hspace{-9cm}
\mbox{\Large EUROPEAN ORGANIZATION FOR NUCLEAR RESEARCH (CERN)}

\vspace*{0.2cm}
\hspace*{-9cm}
\begin{tabular*}{16cm}{lc@{\extracolsep{\fill}}r}
\ifthenelse{\boolean{pdflatex}}
{\vspace*{-2.7cm}\mbox{\!\!\!\includegraphics[width=.14\textwidth]{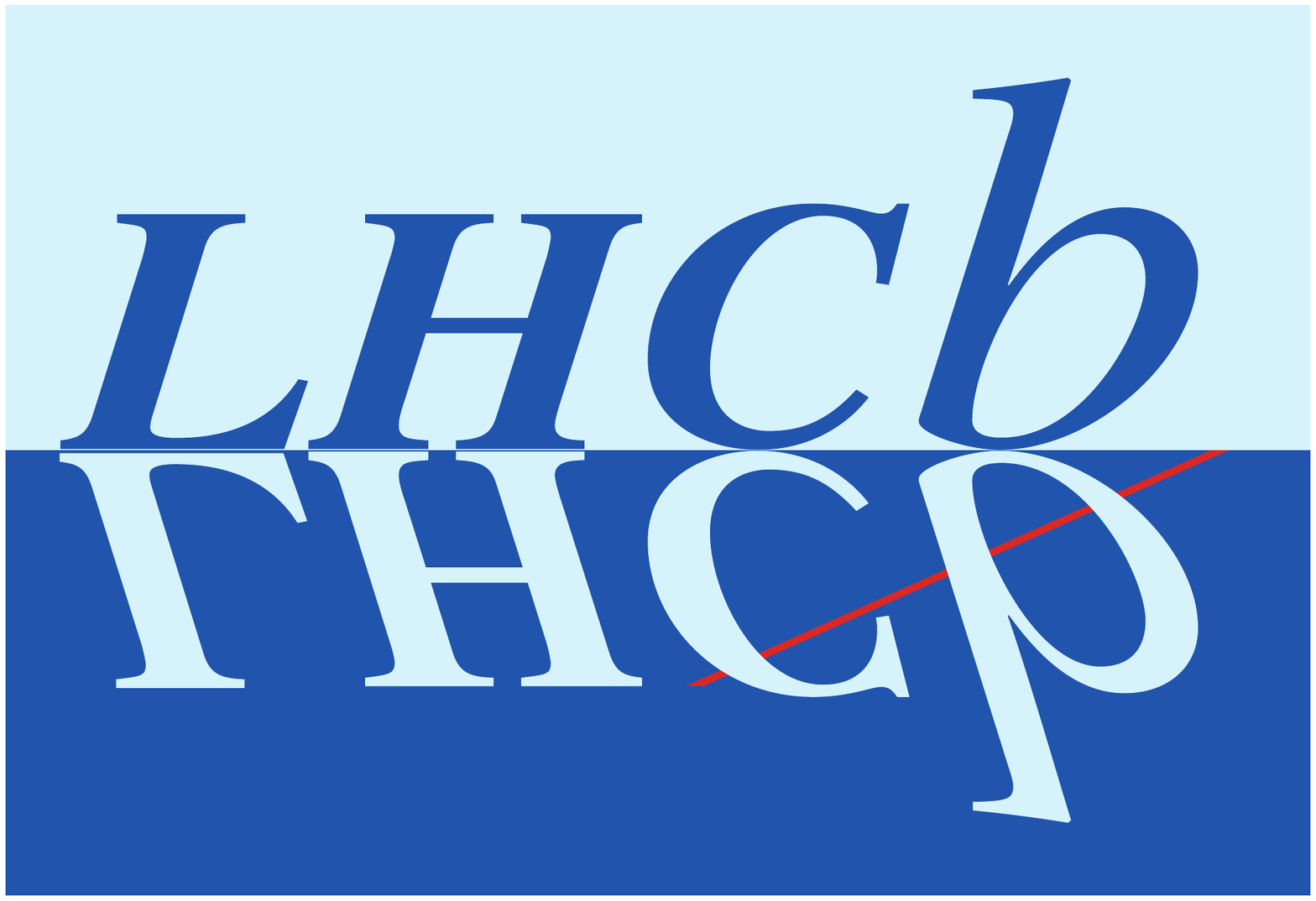}} & &}%
{\vspace*{-1.2cm}\mbox{\!\!\!\includegraphics[width=.12\textwidth]{lhcb-logo.eps}} & &}%
\\
 & & LHCb-PAPER-2011-017 \\
 & & CERN-PH-EP-2011-163 \\ 
 & & \today \\ 
\end{tabular*}
\vspace*{1cm}

\maketitle


\section{Introduction}
\label{sec:Introduction}

To date \CP violation (CPV) has been observed only in decays of neutral
$K$ and $B$ mesons. All observations are consistent with CPV being
generated by the phase in the Cabibbo-Kobayashi-Maskawa (CKM)
matrix~\cite{Cabibbo:1963yz, Kobayashi:1973fv} of the Standard Model (SM).
In the charm sector, CKM dynamics can produce direct \CP asymmetries
in Cabibbo suppressed $D^{\pm}$ decays of the order of 10$^{-3}$ or less~\cite{Bianco:2003vb}. Asymmetries of up to around 1\% can be generated by new
physics (NP)~\cite{Artuso:2008vf, Grossman:2006jg}. In most
extensions of the SM, asymmetries arise in processes with loop
diagrams. However, in some cases CPV could occur even at tree level, for
example in models with charged Higgs exchange. 

In decays of hadrons, CPV can be observed when two different amplitudes
with non-zero relative weak and strong phases contribute coherently to
a final state. Three-body decays are dominated
by intermediate resonant states, and the requirement of a non-zero
relative strong phase is
fulfilled by the phases of the resonances. In two-body decays, CPV leads to
an asymmetry in the partial widths. In three-body
decays, the interference between resonances in the two-dimensional
phase space can lead to observable asymmetries which vary across the
Dalitz plot.

\CP-violating phase differences of $10^{\circ}$ or less do not, in
general, lead to large asymmetries in integrated decay rates, but they
could have clear signatures in the Dalitz plot, as
we will show in Sect.~\ref{sec:toys}. This means that a
two-dimensional search should have higher sensitivity
than an integrated measurement. In addition, the distribution of an
asymmetry across phase space could hint at the underlying dynamics. 
 
At present, no theoretical tools for computing decay fractions and relative phases of
resonant modes in $D$ decays have been applied to multibody \Dp decay modes, and no predictions have been made
for how asymmetries might vary across their Dalitz plots. A full
Dalitz plot analysis of large data samples could, in principle, measure small
phase differences. However, rigorous control of the much larger strong phases would be required. For this to be
achieved, better understanding of the amplitudes,
especially in the scalar sector, will be needed, and effects like three-body final state interactions should be taken into account. 

This paper describes a model-independent search for direct CPV in
the Cabibbo suppressed decay $\Dp \to \Km\Kp\pip$ in a binned
Dalitz plot.\footnote{Throughout
this paper charge conjugation is implied, unless otherwise stated.}
A direct comparison between the $\Dp$ and the $\Dm$ Dalitz plots is made
on a bin-by-bin basis. 
The data sample used is approximately 35~pb$^{-1}$ collected in 2010 by the \lhcb experiment at a centre of mass
energy of $\sqrt{s} = 7$ TeV.
This data set corresponds to nearly 10 and 20 times more signal events
than used in 
previous studies of this channel performed by the \babar~\cite{Aubert:2005gj} and 
\mbox{CLEO-c~\cite{:2008zi}} collaborations, respectively. It is
comparable to the dataset used in a more recent search for CPV in $\Dp \to \phi\pip$ decays at \belle~\cite{:2011en}.

The strategy is as follows. For each bin in the Dalitz plot, a local \CP 
asymmetry variable is defined~\cite{Bediaga:2009tr, Aubert:2008yd},

\begin{equation}
\mathcal{S}_{CP}^i = \frac{N^i(\Dp)-\alpha N^i(\Dm)}
{\sqrt{N^i(\Dp)+\alpha^2 N^i(\Dm)}} \ , \hskip .5cm  
\alpha=\frac{N_{\mathrm{tot}}(\Dp)}{N_{\mathrm{tot}}(\Dm)},
\label{intro:signif}
\end{equation}
where $N^i(\Dp)$ and $N^i(\Dm)$ are the numbers of $D^{\pm}$ candidates in the $i^{\mathrm{th}}$ bin and
$\alpha$ is the ratio between the total $\Dp$ and $\Dm$ yields. The
parameter $\alpha$ accounts for global asymmetries, i.e. those that are constant across
the Dalitz plot.

In the absence of Dalitz plot dependent asymmetries, the  $\mathcal{S}_{CP}^i$ values are distributed 
according to a Gaussian distribution with zero mean and unit width. CPV signals are, therefore,
deviations from this behaviour. The numerical comparison between the $\Dp$ and the $\Dm$
Dalitz plots is made with a $\chi^2$ test ($\chi^2=\sum (\mathcal{S}_{CP}^i)^2$). The number 
of degrees of freedom is the number of bins minus one (due to the constraint of the
overall $\Dp/\Dm$ normalization). The $p$-value that results from this
test is defined as the
probability of obtaining, for a given number of degrees of freedom and
under the assumption of no CPV, a
$\chi^{2}$ that is at least as high as the value observed \cite{lyons1989statistics}. It measures the degree to which we are confident that
the differences between the $\Dp$ and $\Dm$ Dalitz plots are driven
only by statistical fluctuations.

If CPV is observed, the $p$-value from this test could be converted into a
significance for a signal using Gaussian statistics. However, in the
event that no CPV is found, there is no model-independent mechanism for setting limits
on CPV within this procedure. In this case, the results
can be compared to simulation studies in which an artificial \CP
asymmetry is introduced into an assumed amplitude model for the
decay. Since such simulations are clearly
model-dependent, they are only used as a guide to
the sensitivity of the method, and not in the determination of the $p$-values that constitute
the results of the analysis.

The technique relies on careful accounting for local asymmetries that could be induced  
by sources such as, the difference 
in the $K$--nucleon inelastic cross-section, differences in the
reconstruction or trigger efficiencies, 
left-right detector asymmetries, etc. These effects are investigated
in the two control channels $\Dp \to \Km\pip\pip$ and
$\Dsp \to \Km\Kp\pip$.

The optimum sensitivity is obtained with bins of nearly the same size
as the area over which the asymmetry extends in the Dalitz plot. Since this is a search for new and therefore unknown phenomena, it is
necessary to be sensitive to effects restricted to small areas as well
as those that extend over a large region of the Dalitz plot. Therefore two types of binning scheme
are employed. The first type is simply a uniform
grid of equally sized bins. The second type takes into account the fact
that the 
$\Dp \to \Km\Kp\pip$ Dalitz plot is dominated by the $\phi\pi^+$ and $\Kbar^*(892)^{0}K^+$
resonances, so the event distribution is highly non-uniform. This ``adaptive binning'' scheme uses 
smaller bins where the density of events is high, aiming for a uniform
bin population. In each scheme, different numbers of bins are used 
in our search for localized asymmetries. 

The paper is organized as follows. In Sect.~\ref{sec:data}, a description of the \lhcb experiment 
and of the data selection is presented. In Sect.~\ref{sec:toys}, the methods and the binnings are
discussed in detail. The study of the control channels and of possible asymmetries
generated by
detector effects or backgrounds is presented in Sect.~\ref{sec:control}. The results of our search are given in Sect.~\ref{sec:results}, and the
conclusions in Sect.~\ref{sec:conclusion}.

\section{Detector, dataset and selection}
\label{sec:data}

The \lhcb detector~\cite{Alves:2008zz} is a single-arm forward spectrometer with the main purpose of measuring CPV and rare decays of hadrons containing $\Pb$ and $\Pc$ quarks.  A vertex locator (\velo) determines with high precision the positions of the vertices of primary $\proton\proton$ collisions (PVs) and the decay vertices of long-lived particles. The tracking system also includes a large area  silicon strip detector located in front
of a dipole magnet with an integrated field of around  4~Tm, and a combination of silicon strip detectors and straw drift chambers placed behind the magnet.  Charged hadron identification is achieved with two ring-imaging Cherenkov (RICH) detectors. The calorimeter system consists of a preshower, a scintillator pad
detector, an electromagnetic calorimeter and a hadronic calorimeter. It identifies high
transverse energy ($\et$) hadron, electron and photon candidates and provides information
for the trigger. Five muon stations composed of multi-wire proportional chambers and triple-GEMs (gas electron multipliers) provide fast information for the trigger and muon
identification capability.

The \lhcb trigger consists of two levels. The first, hardware-based level selects leptonic and hadronic final states with high tranverse momentum, using the subset of the detectors that are able to reduce the rate at which the whole detector is read out to a maximum of 1~MHz. 
The second level, the High Level Trigger (HLT), is subdivided into two software stages that can use the information from all parts of the detector. The first stage, HLT1, performs a partial reconstruction of the event, reducing the rate further and allowing the next stage, HLT2, to fully reconstruct the individual channels. At each stage, several selections designed for specific types of decay exist.  As luminosity increased throughout 2010 several changes in the trigger were required.  To match these, the datasets for signal and control modes are divided into three parts according to the trigger, samples 1, 2 and 3, which correspond to integrated luminosities of approximately 3, 5 and 28~pb$^{-1}$ respectively. The magnet polarity was changed several times during data taking.

The majority of the signal decays come via the hadronic hardware trigger, which has an $\et$ threshold that varied between 2.6 and 3.6~GeV in 2010. In the HLT1, most candidates also come from the hadronic selections which retain events with at least one high transverse momentum ($\pt$) track that is displaced from the PV. In the HLT2, dedicated charm triggers select most of the signal. However, the signal yield for these channels can be increased by using other trigger selections, such as those for decays of the form $B \to DX$. To maintain the necessary control of Dalitz plot-dependent asymmetries, only events from selections which have been measured not to introduce charge asymmetries into the Dalitz plot of the $\Dp \to \Km\pip\pip$ control mode are accepted.

The signal ( $\Dp \to \Km\Kp\pip$) and control ($\Dp \to \Km\pip\pip$ and $\Dsp \to \Km\Kp\pip$) mode candidates are selected using the same criteria, which are chosen to maximize the statistical significance of the signal. Moreover, care is taken to use selection cuts that do not have a low efficiency in any part of the Dalitz plot, as this would reduce the sensitivity in these areas. 
The  selection criteria are the same regardless of the trigger conditions.

The event selection starts by requiring at least one PV with a minimum of five charged tracks to exist. To control CPU consumption each event must also have fewer than 350 reconstructed tracks. The particle identification system constructs a relative log-likelihood for pion and kaon hypotheses, \dllkpi, and we require \dllkpi $>$ 7 for kaons and $<$ 3 for pions.  
Three particles with appropriate charges are combined to form the $D^+_{(s)}$ candidates. 
The corresponding tracks are required to have a  good fit quality  ($\chi^2/{\rm ndf} < 5$), $\pt >$ 250~MeV$/c$, momentum $p >$ 2000~MeV$/c$ and the scalar sum of their $\pt$ above 2800~MeV$/c$. 
Because a typical \Dp travels for around 8~mm before decaying, the final state tracks should not point to the PV. 
The smallest displacement from each track to the PV is computed, and a $\chi^2$ ($\chi^{2}_{\mathrm{IP}}$), formed
by using the hypothesis that this distance is equal to zero, is required to be greater than 4 for each track.
The three daughters should be produced at a common origin, the
charm decay vertex, with vertex fit $\chi^2/{\rm ndf} <$ 10.

This `secondary' vertex must be well separated from any PV, thus a flight distance variable ($\chi^2_{\mathrm{FD}}$) is constructed. The secondary vertex is required to have $\chi^2_{\mathrm{FD}} > 100$, and to be downstream of the PV. The \pt of the $D^{+}_{(s)}$ candidate must be greater than 1000~MeV$/c$, and its reconstructed trajectory is required to originate from the PV ($\chi^2_{\mathrm{IP}} < 12$). 

\begin{figure*}[tbp]
  \begin{center}

    \includegraphics*[width=0.48\textwidth]{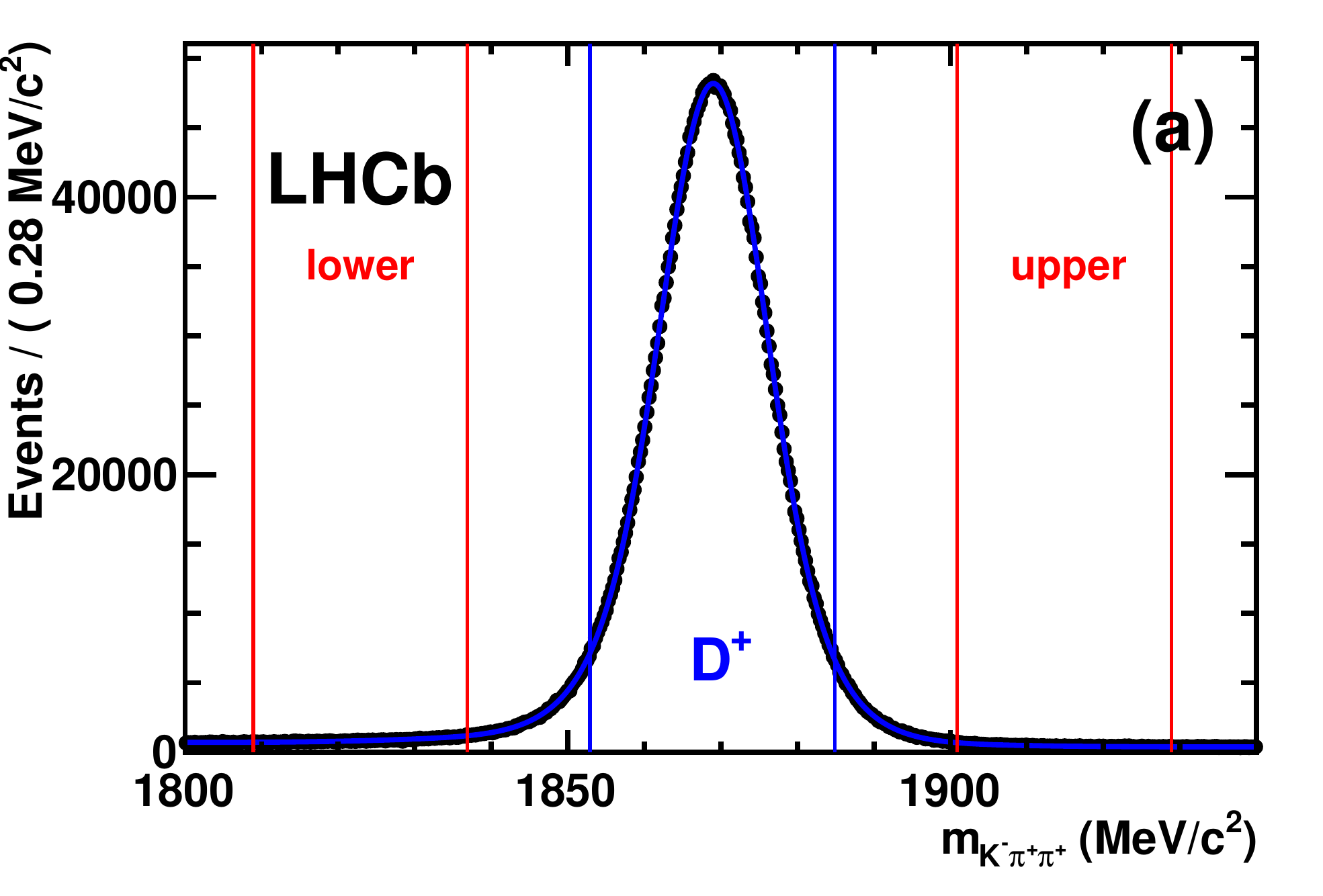}
    \includegraphics*[width=0.48\textwidth]{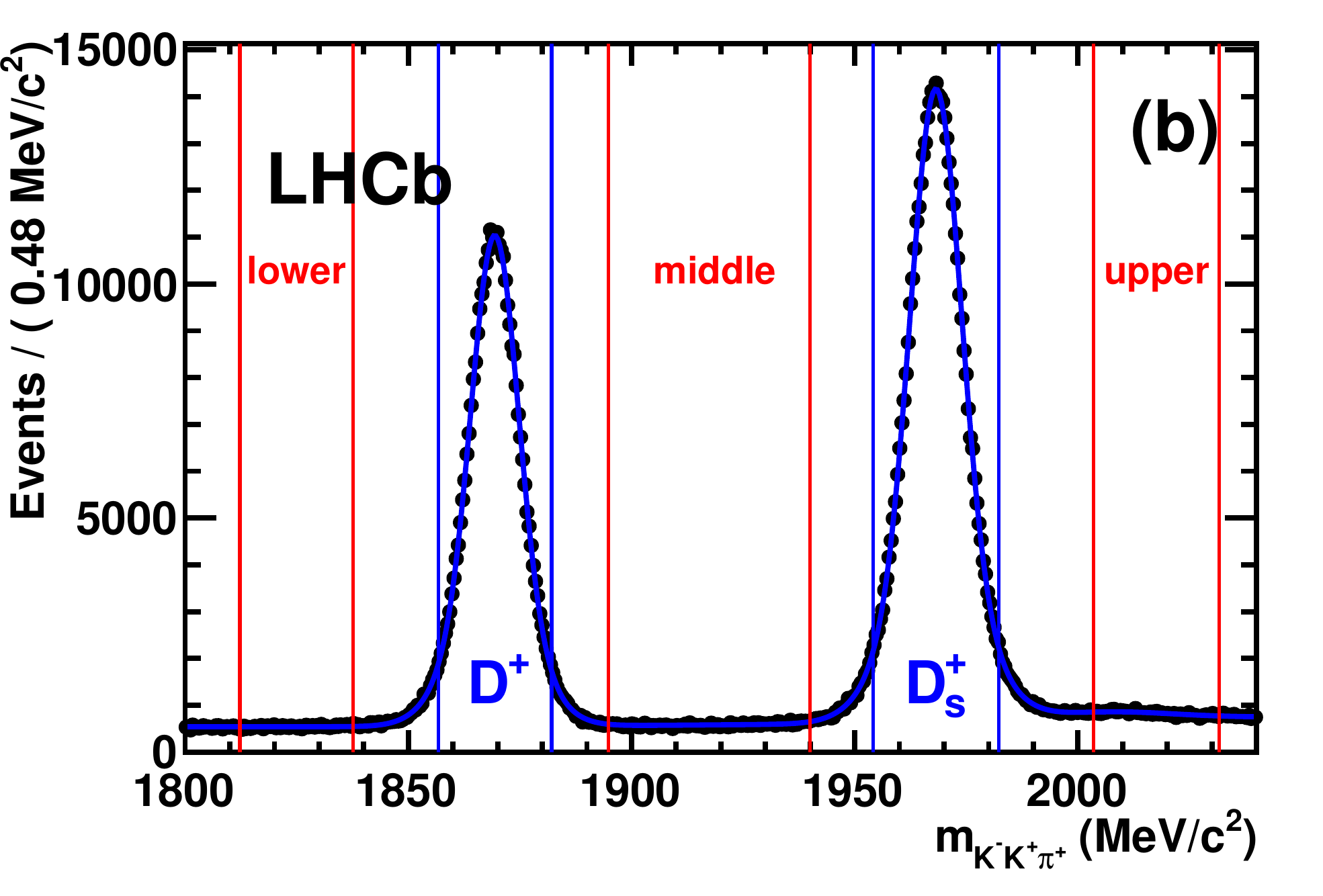}

  \end{center}
  \caption{Fitted mass spectra of (a) $\Km\pip\pip$ and (b) $\Km\Kp\pip$ candidates from samples 1 and 3, $\Dp$ and $\Dm$ combined. The signal mass windows and sidebands defined in the text are labelled. }
  \label{fig:data:massfit}
\end{figure*}

\begin{figure}[tbp]
  \begin{center}

    \includegraphics*[width=0.5\textwidth]{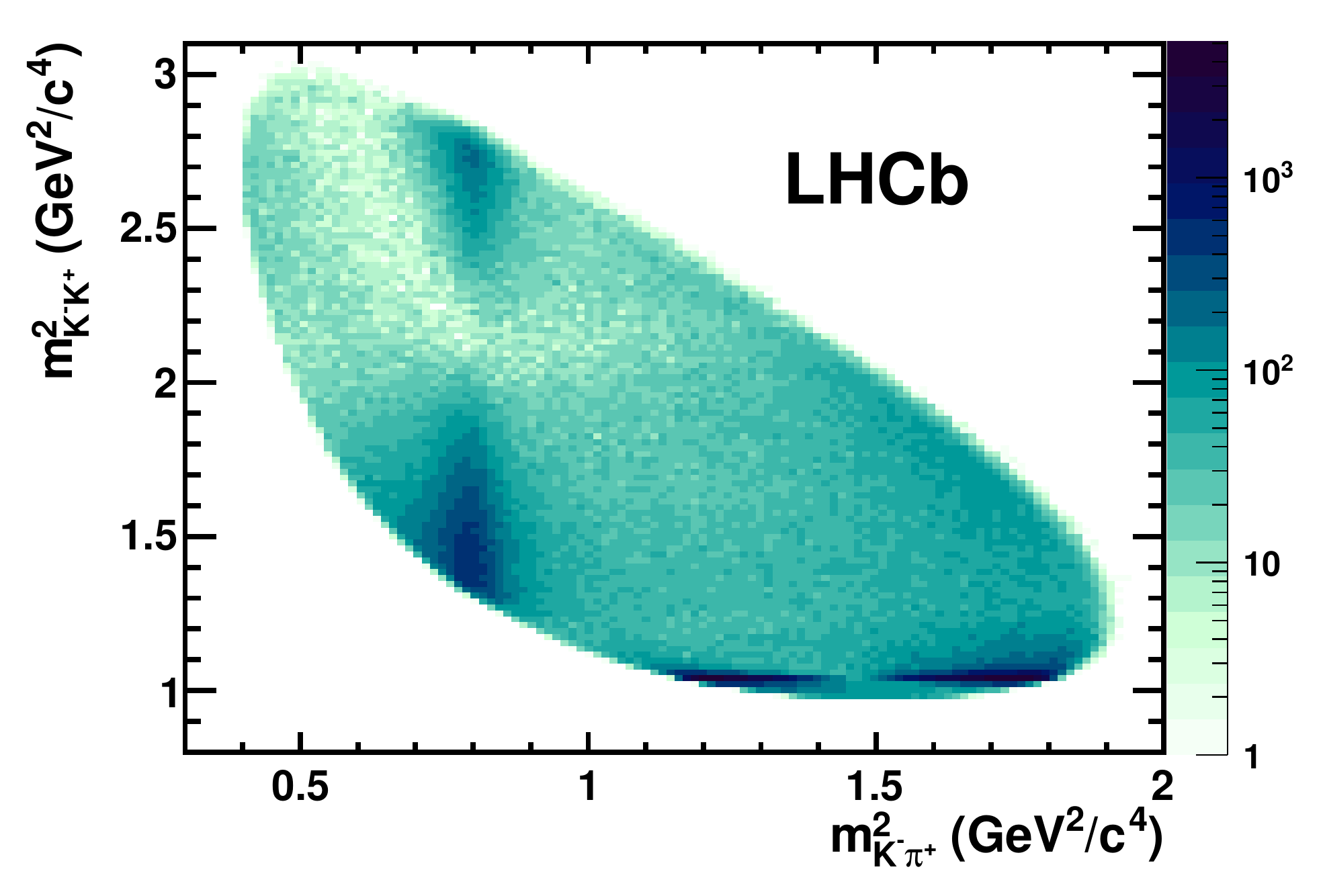}

  \end{center}
  \caption{Dalitz plot of the $\Dp \to \Km\Kp\pip$ decay for selected candidates in the signal window. The vertical $\Kbar^{*}(892)^{0}$ and horizontal $\phi(1020)$ contributions are clearly visible in the data. }
  \label{fig:data:dalitz}
\end{figure}

In order to quantify the signal yields ($S$), a simultaneous fit to the invariant mass distribution of the \Dp and \Dm samples is performed. A double Gaussian is used for the $\Km\Kp\pip$ signal, whilst the background ($B$) is described by a quadratic component and a single Gaussian for the small contamination from $\Dstarp \to \Dz (\Km\Kp)\pip$ above the \Dsp peak. The fitted mass spectrum for samples 1 and 3 combined is shown in Fig.~\ref{fig:data:massfit}, giving the yields shown in 
Table~\ref{table:data:pur}.  A weighted mean of the widths of the two Gaussian contributions to the mass peaks is used to determine the overall widths, $\sigma$, as 6.35~MeV/$c^2$ for $\Dp \to \Km\Kp\pip$, 7.05~MeV/$c^2$ for $\Dsp \to \Km\Kp\pip$, and 8.0~MeV/$c^2$ for $\Dp \to \Km\pip\pip$.  These values are used to define signal mass windows of approximately 2$\sigma$ in which the Dalitz plots are constructed. The purities, defined as $S/(B+S)$ within these mass regions, are also shown in Table~\ref{table:data:pur} for samples 1 and 3 in the different decay modes. 

For sample 2, the yield cannot be taken directly from the fit, because there is a mass cut in the HLT2 line that accepts the majority of the signal, selecting events in a $\pm25$~MeV$/c^2$ window around the nominal value. However, another HLT2 line with a looser mass cut that is otherwise identical to the main HLT2 line exists, although only one event in 100 is retained. In this line the purity is found to be the same in sample 2 as in sample 3. The yield in sample 2 is then inferred as the total $(S+B)$ in all allowed triggers in the mass window times the purity in sample 3. Thus the overall yield of signal $\Dp \to \Km\Kp\pip$ candidates in the three samples within the mass window is approximately 370,000. The total number of candidates ($S+B$) in each decay mode used in the analysis are given in Table~\ref{table:data:cand}. The Dalitz plot of data in the \Dp window is shown in Fig.~\ref{fig:data:dalitz}.

\begin{table}[tbp]

\caption{
Yield ($S$) and purity for samples 1 and 3 after the final selection. The purity is estimated in the 2$\sigma$ mass window.
 \label{table:data:pur}}

\begin{center}
  \begin{tabular}{cccc} \hline\hline
Decay & Yield &  \multicolumn{2}{c}{Purity} \\
 & Sample 1+3& Sample 1& Sample 3\\\hline
$\Dp \to \Km\Kp\pip$   & $(3.284\pm 0.006) \times 10^5$& 88\%  & 92\%  \\
$\Dsp\to \Km\Kp\pip$ & $(4.615 \pm 0.012)\times 10^5$ & 89\%  & 92\%  \\
$\Dp\to \Km\pip\pip$ & $(3.3777 \pm 0.0037)\times 10^6$ & 98\%  &  98\%  \\
\hline\hline
\end{tabular}
  \end{center}

\end{table}

\begin{table}[tbp]
    \caption{Number of candidates $(S+B)$ in the signal windows shown in Fig.~\ref{fig:data:massfit} after the final selection, for use in the subsequent analysis.
  \label{table:data:cand}}

 \begin{center}
    \begin{tabular}{ccccc}\hline\hline
        &    sample 1 & sample 2 & sample 3 & Total\\\hline
$\Dp\to \Km\Kp\pip$  & 84,667 & 65,781 & 253,446 & 403,894  \\
$\Dsp \to \Km\Kp\pip $ &   126,206 & 91,664 & 346,068 & 563,938  \\
$\Dp \to \Km\pip\pip$ &    858,356 & 687,197 & 2,294,315 & 3,839,868\\
\hline\hline
    \end{tabular}
  \end{center}
\end{table}

\begin{table*}[tbp]
\caption[CLEO fit]{
The CLEO-c amplitude model ``B'' \cite{:2008zi} used in the simulation studies. The uncertainties
are statistical, experimental systematic and model systematic
respectively. 
 \label{table:toys:cleoB}}

\begin{center}
  \begin{tabular*}{\linewidth}{@{\extracolsep{\fill}}llll} \hline\hline
Resonance &  $\phantom{000}$ Amplitude & $\phantom{.}$ Relative phase & $\phantom{0}$ Fit fraction\\
\hline
$\Kbar^{*}(892)^{0}$ &$ \phantom{000}$ 1 (fixed) 
& $\phantom{000}$ 0 (fixed) & $25.7\pm0.5^{+0.4+0.1}_{-0.3-1.2}$\\
$\Kbar^{*}_{0}(1430)^{0}$ &
$\phantom{0}4.56\pm0.13^{+0.10+0.42}_{-0.01-0.39}$ 
& $\phantom{-}\phantom{0}70\pm6^{+1+16}_{-6-23}$ & $18.8\pm1.2^{+0.6+3.2}_{-0.1-3.4}$ \\
$\kappa(800)$& $\phantom{0}2.30\pm0.13^{+0.01+0.52}_{-0.11-0.29}$ &
$\phantom{}-87\pm6^{+2+15}_{-3-10}$  
&$\phantom{0}7.0\pm0.8^{+0.0+3.5}_{-0.6-1.9}$ \\
$\Kbar^{*}_{2}(1430)^{0}$ &
$\phantom{00}7.6\pm0.8^{+0.5+2.4}_{-0.6-4.8}$ 
& $\phantom{-}171\pm4^{+0+24}_{-2-11}$ & $\phantom{0}1.7\pm0.4^{+0.3+1.2}_{-0.2-0.7}$ \\
$\phi(1020)$ & $1.166\pm0.015^{+0.001+0.025}_{-0.009-0.009}$ &
$-163\pm3^{+1+14}_{-1-5}$ & $27.8\pm0.4^{+0.1+0.2}_{-0.3-0.4}$ \\
$a_{0}(1450)^{0}$ & $\phantom{0}1.50\pm0.10^{+0.09+0.92}_{-0.06-0.33}$ 
& $\phantom{-}116\pm2^{+1+7}_{-1-14}$ & $\phantom{0}4.6\pm0.6^{+0.5+7.2}_{-0.3-1.8}$ \\
$\phi(1680)$& $\phantom{0}1.86\pm0.20^{+0.02+0.62}_{-0.08-0.77}$ 
& $-112\pm6^{+3+19}_{-4-12}$ & $0.51\pm0.11^{+0.01+0.37}_{-0.04-0.15}$ \\

\hline\hline
\end{tabular*}
\end{center}
\end{table*}

Within the $2\sigma$ $\Dp \to \Km\Kp\pip$ mass window, about 8.6\% of events are background.  Apart from random three-body track combinations, charm backgrounds and two-body resonances plus one track are expected. Charm reflections appear when a particle is wrongly identified in a true charm three-body decay and/or a track in a four-body charm decay is lost. The main three-body reflection in the $\Km\Kp\pip$ spectrum is the Cabibbo-favoured $\Dp \to \Km\pip\pip$, where the incorrect assignment of the kaon mass to the pion leads to a distribution that partially overlaps with the $\Dsp \to \Km\Kp\pip$ signal region, but not with $\Dp \to \Km\Kp\pip$.  The four body, Cabibbo-favoured mode $\Dz \to \Km\pip\pim\pip$ where a $\pi^+$ is lost and the $\pi^-$ is misidentified as a $K^-$ will appear broadly distributed in $\Km\Kp\pip$ mass, but its resonances could create structures in the Dalitz plot. Similarly, $\Kbar^*(892)^0$ and $\phi$ resonances from the PV misreconstructed with a random track forming a three-body vertex will also appear.

\section{Methods and binnings}
\label{sec:toys}

\begin{figure*}[tbp]
  \begin{center}
\includegraphics*[width=0.45\textwidth]{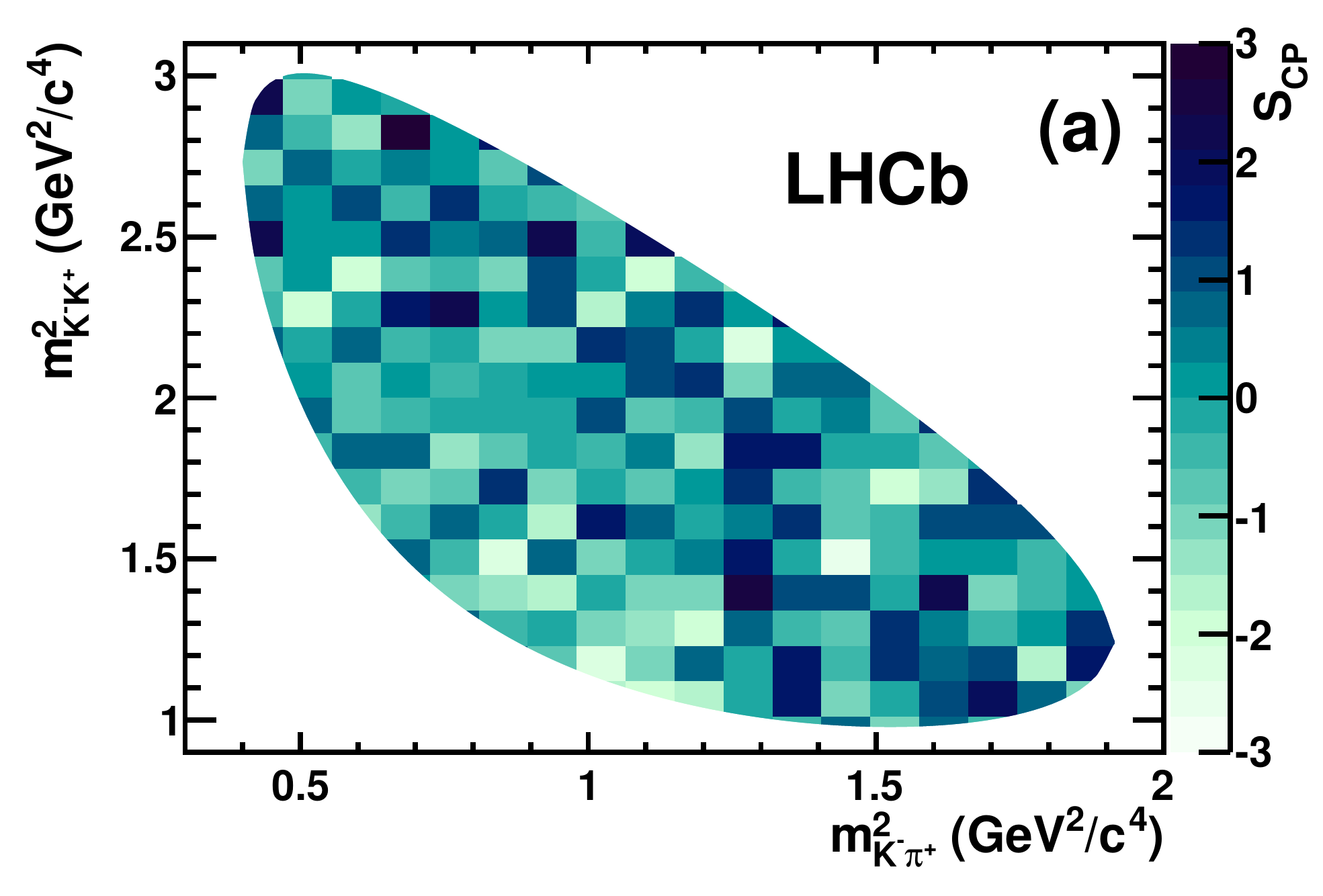}
    \includegraphics*[width=0.45\textwidth]{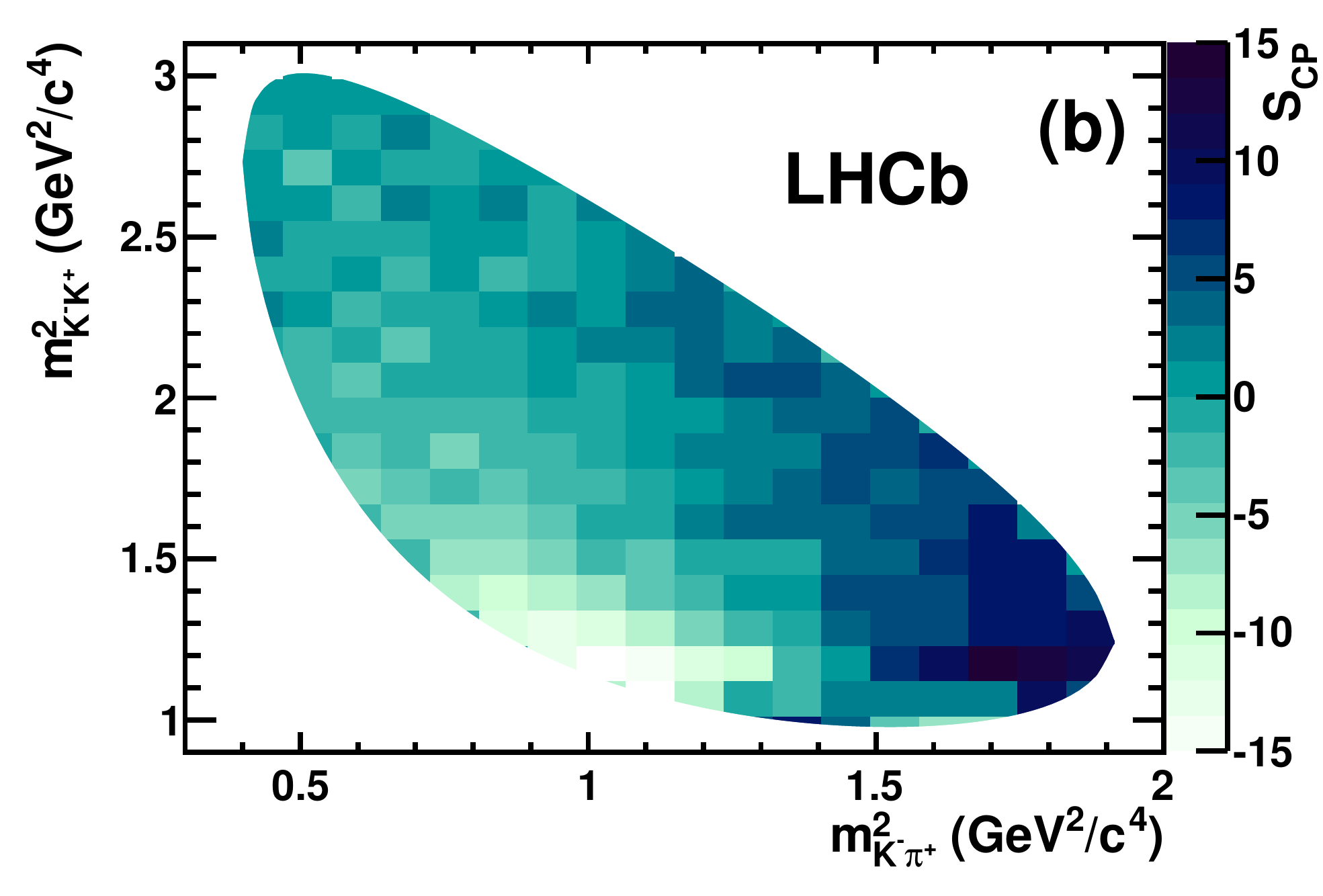}
  \end{center}
  \caption{$\mathcal{S}_{CP}$ across the Dalitz plot in a Monte Carlo pseudo-experiment with a large number of events with (a) no CPV and (b) a 4$^{\circ}$ CPV in the $\phi\pi$ phase. Note the difference in colour scale between (a) and (b).}
  \label{fig:highstats}
\end{figure*}
\begin{figure*}[tbp]
\begin{center}
\includegraphics*[width=0.45\textwidth]{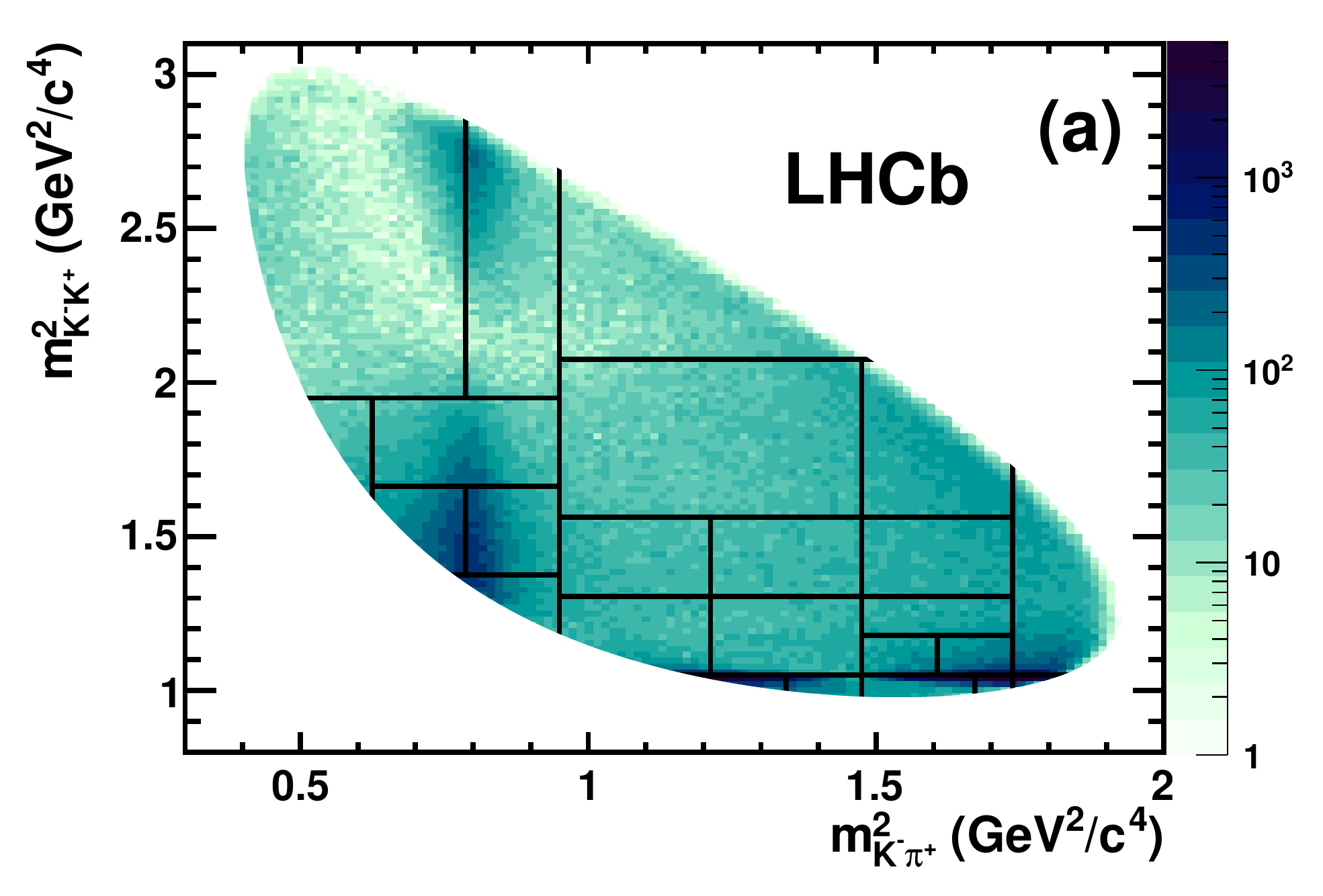}
\includegraphics*[width=0.45\textwidth]{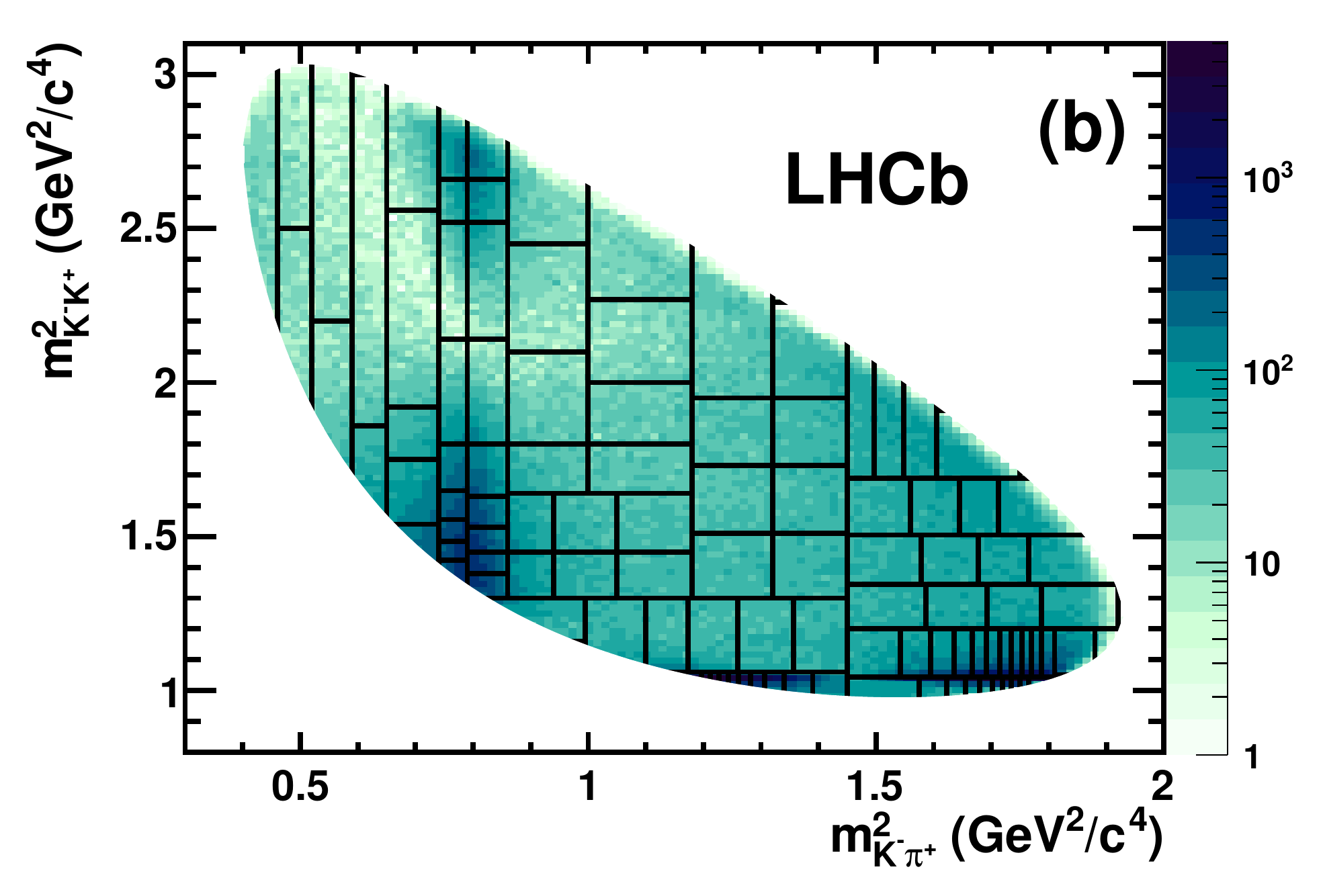}
 \end{center}
 \caption{ Layout of the (a) ``Adaptive I''  and (b) ``Adaptive II'' binnings on the Dalitz plot of data.}
 \label{fig:toydist1}
\end{figure*}

Monte Carlo pseudo-experiments are used to verify that we can detect CPV with the strategy outlined in Sect. \ref{sec:Introduction} without producing fake signals, and to devise and test suitable binning schemes for the Dalitz plot. They are also used to quantify our sensitivity to possible manifestations of CPV, where we define the sensitivity to a given level of CPV as the probability of observing it with $3\sigma$ significance.

 For the $D^+ \to K^-K^+\pi^+$ Dalitz plot model, the result of the \mbox{CLEO-c} analysis (fit B) \cite{:2008zi} is used. The amplitudes and phases of the resonances used in this model are reproduced in Table~\ref{table:toys:cleoB}.  For simplicity, only resonant modes with fit fractions greater than $2\%$ are included in the pseudo-experiments. The fit fraction for a resonance is defined as the
integral of its squared amplitude over the Dalitz plot divided by the integral
of the square of the overall complex amplitude. An efficiency function is determined from a two-dimensional second order polynomial fit to the Dalitz plot distribution of triggered events that survive the selection cuts in the GEANT-based  \cite{Agostinelli:2002hh} LHCb Monte Carlo simulation for nonresonant  $\Dp \to \Km \Kp \pip$. A simple model for the background is inferred from the Dalitz plots of the sidebands of the $\Dp \to \Km \Kp \pip$ signal.  It is composed of random combinations of $\Km$, $\Kp$, and $\pip$ tracks,  $\phi$ resonances with $\pip$ tracks, and $\Kbar^{*}(892)^{0}$ resonances with $\Kp$ tracks. The \mbox{CLEO-c} Dalitz plot analysis has large uncertainties, as do the background and efficiency simulations (due to limited numbers of MC events), so the method is tested on a range of different Dalitz plot models.

Pseudo-experiments with large numbers of events are used to investigate how CPV would be observed in the Dalitz plot. These experiments are simple ``toy'' simulations that produce points in the Dalitz plot according to the probability density function determined from the \mbox{CLEO-c} amplitude model with no representation of the proton-proton collision, detector, or trigger. Figure~\ref{fig:highstats}(a) illustrates the values of $\mathcal{S}_{CP}^i$ observed with $8\times10^{7}$ events and no CPV. This dataset is approximately 50 times larger than the data sample under study. The resulting $\chi^{2}/{\rm ndf}$ is $253.4/218$, giving a $p$-value for consistency with no CPV of 5.0\%. This test shows that the method by itself is very unlikely to yield false positive results. Figure~\ref{fig:highstats}(b) shows an example test using $5\times10^{7}$ events with a \CP violating phase difference of $4^{\circ}$ between the amplitudes for the $\phi(1020)\pip$ component in $\Dp$ and $\Dm$ decays. The $p$-value in this case is less than $10^{-100}$.
 The CPV effect is clearly visible, and is spread over a broad area of the plot, changing sign from left to right. This sign change means the CPV causes only a 0.1\% difference in the total decay rate between $\Dp$ and $\Dm$. This illustrates the strength of our method, as the asymmetry would be much more difficult to detect in a measurement that was integrated over the Dalitz plot. Even with no systematic uncertainties, to see a 0.1\% asymmetry at the $3\sigma$ level would require $2.25\times10^{6}$ events. With the method and much smaller dataset used here we would observe this signal at the $3\sigma$ level with 76\% probability, as shown in Table~\ref{table:toyresults} below.

The sensitivity to a particular manifestation of CPV depends on the choice of binning. The fact that the \CP-violating region in most of the pseudo-experiments covers a broad area of the Dalitz plot suggests that the optimal number of bins for this type of asymmetry is low. Each bin adds a degree of freedom without changing the $\chi^{2}$ value for consistency with no CPV. However, if \CP asymmetries change sign within a bin, they will not be seen. Similarly, the sensitivity is reduced if only a small part of a large bin has any CPV in it. To avoid effects due to excessive fluctuations, bins that contain fewer than 50 candidates are not used anywhere in the analysis. Such bins are very rare.

The binnings are chosen to reflect the highly non-uniform structure of the Dalitz plot. A simple adaptive binning algorithm was devised to define binnings of approximately equal population without separating $D^+$ and $D^-$. Two binnings that are found to have good sensitivity to the simulated asymmetries contain 25 bins (``Adaptive I'') arranged as shown in Fig.~\ref{fig:toydist1}(a), and 106 bins (``Adaptive II'') arranged as shown in Fig.~\ref{fig:toydist1}(b).  For Adaptive I,  a simulation of the relative value of the strong phase across the Dalitz plot in the \mbox{CLEO-c} amplitude model is used to refine the results of the algorithm: if the strong phase varies significantly across a bin, \CP asymmetries are more likely to change sign. Therefore the bin boundaries are adjusted to minimise changes in the strong phase within bins. The model-dependence of this simulation could, in principle, influence the binning and therefore the sensitivity to CPV, but it cannot introduce model-dependence into the final results as no artificial signal could result purely from the choice of binning. Two further binning schemes, ``Uniform I'' and ``Uniform II'', are defined. These use regular arrays of rectangular bins of equal size.
\begin{table*}[tbp]
\caption[Toy results]{
Results from sets of 100 pseudo-experiments with different \CP asymmetries and Adaptive I and II binnings. $p(3\sigma)$ is the probability of a 3$\sigma$
observation of CPV. $\langle S \rangle$ is the mean significance with which CPV
is observed. 
\label{table:toyresults}}
\begin{center}
\setlength{\tabcolsep}{10pt}
  \begin{tabular}{ccccc}
    \hline\hline
CPV & \multicolumn{2}{c}{Adaptive I} & \multicolumn{2}{c}{Adaptive II}\\
     & $p(3\sigma)$& $\langle S \rangle$  &
    $p(3\sigma)$ & $\langle S \rangle$ \\
\hline
no CPV &0 &0.84$\sigma$ & 1\% & 0.84$\sigma$ \\ 
$6^{\circ}$ in $\phi(1020)$ phase  & 99\%& 7.0$\sigma$ &98\% &5.2$\sigma$ \\ 
$5^{\circ}$ in $\phi(1020)$ phase &97\% & 5.5$\sigma$& 79\% &3.8$\sigma$ \\ 
$4^{\circ}$ in $\phi(1020)$ phase  &76\% &3.8$\sigma$ & 41\% & 2.7$\sigma$\\ 
$3^{\circ}$ in $\phi(1020)$ phase  &38\%  &2.8$\sigma$ &12\% & 1.9$\sigma$\\
$2^{\circ}$ in $\phi(1020)$ phase &5\% & 1.6$\sigma$  & 2\% &1.2$\sigma$\\
$6.3\%$ in $\kappa(800)$ magnitude & 16\% & 1.9$\sigma$ & 24\% & 2.2$\sigma$ \\
$11\%$ in $\kappa(800)$ magnitude & 83\% & 4.2$\sigma$ & 95\% & 5.6$\sigma$ \\ 
\hline\hline
\end{tabular}
\end{center}
\end{table*}

\begin{table*}[tbp]
\caption[Toy results]{
Results from sets of 100 pseudo-experiments with $4^{\circ}$ CPV in the
$\phi(1020)$ phase
and different Dalitz plot models. $p(3\sigma)$ is the probability of a 3$\sigma$
observation of CPV. $\langle S \rangle$ is the mean significance with which CPV
is observed.  The sample size is comparable to that seen in data.
\label{tab:varymodel}}

\begin{center}
\setlength{\tabcolsep}{10pt}
  \begin{tabular}{ccccc} \hline\hline
Model & \multicolumn{2}{c}{Adaptive I} & \multicolumn{2}{c}{Adaptive II}\\
     & $p(3\sigma)$& $\langle S \rangle$  &
    $p(3\sigma)$ & $\langle S \rangle$ \\
\hline
B (baseline) &76\% & 3.8$\sigma$& 41\% & 2.7$\sigma$\\  
A  & 84\% & 4.3$\sigma$ &47\% &2.9$\sigma$ \\  
B2 (add $f_{0}(980)$) &53\% &3.2$\sigma$ & 24\% &2.2$\sigma$ \\ 
B3 (vary $\Kbar^{*}_{0}(1430)^0$ magn.)  & 82\% & 4.0$\sigma$ & 41\%& 2.8$\sigma$\\ 
B4 (vary $\Kbar^{*}_{0}(1430)^0$ phase)  & 73\% & 3.7$\sigma$ &38\% & 2.7$\sigma$\\ 
\hline\hline
\end{tabular}
\end{center}
\end{table*}

The adaptive binnings are used to determine the sensitivity to several manifestations of CPV. With 200 test experiments of approximately the same size as the signal sample in data, including no asymmetries, no \CP-violating signals are observed at the 3$\sigma$ level with Adaptive I or Adaptive II. The expectation is 0.3. 

With the chosen binnings, a number of sets of 100 pseudo-experiments with different \CP-violating asymmetries are produced. The probability of observing a given signal in either the $\phi(1020)$ or $\kappa(800)$ resonances with 3$\sigma$ significance is calculated in samples of the same size as the dataset. The results are given in Table~\ref{table:toyresults}. The CPV shows up both in the $\chi^{2}/{\rm ndf}$ and in the width of the fitted $\mathcal{S}_{CP}$ distribution. 

For comparison, the asymmetries in the $\phi$ phase and $\kappa$ magnitude measured
by the CLEO collaboration using the same amplitude model were $(6\pm6^{+0+6}_{-2-2})^{\circ}$ and
$(-12\pm12^{+6+2}_{-1-10})\%$,\footnote{The conventions used in the CLEO paper to define asymmetry are different, so the asymmetries in Table II of \cite{:2008zi} have been multiplied by two in order to be comparable with those given above.} where the uncertainties are
statistical, systematic and model-dependent, respectively. Table~\ref{table:toyresults} suggests that, assuming their model, we would be at least 95\% confident of detecting the central values of these asymmetries.

The sensitivity of the results to variations in the Dalitz plot model and the background is investigated, and example results for the \CP asymmetry in the $\phi(1020)$ phase are shown in Table~\ref{tab:varymodel}. In this table, models A and B are taken from the CLEO paper, model B2 includes an $f_{0}(980)$ contribution that accounts for approximately 8\% of events, and models B3 and B4 are variations of the $\Kbar^{*}_{0}(1430)^{0}$ amplitude and phase within their uncertainties. As expected, the sensitivity to CPV in the resonances of an amplitude model depends quite strongly on the details of the model. This provides further justification for our model-independent approach. However, a reasonable level of sensitivity is retained in all the cases we tested.  Thus, when taken together, the studies show that the method works well. It does not yield fake signals, and should be sensitive to any large CPV that varies significantly across the Dalitz plot even if it does not occur precisely in the way investigated here.

\section{Control modes}
\label{sec:control}

It is possible that asymmetries exist in the data that do not result
from CPV, for
example due to production, backgrounds, instrumental effects such as left-right differences in detection 
efficiency, or momentum-dependent differences in the interaction cross-sections of the daughter particles with detector material. Our
sensitivity to such asymmetries is
investigated in the two Cabibbo favoured 
control channels, where there is no large CPV predicted. 
The $\Dp \to \Km\pip\pip$ control mode has an order of magnitude
more candidates
than the Cabibbo-suppressed signal mode, and is more
sensitive to detector effects since there is no
cancellation between $\Kp$ and $\Km$ reconstruction
efficiencies. Conversely, the $\Dsp \to \Km\Kp\pip$ 
control mode is very similar to our signal mode in terms of 
resonant structure, number of candidates, kinematics,
detector effects, and backgrounds.

\begin{figure*}[tbp]
\begin{center}
  \includegraphics[width=0.45\textwidth]{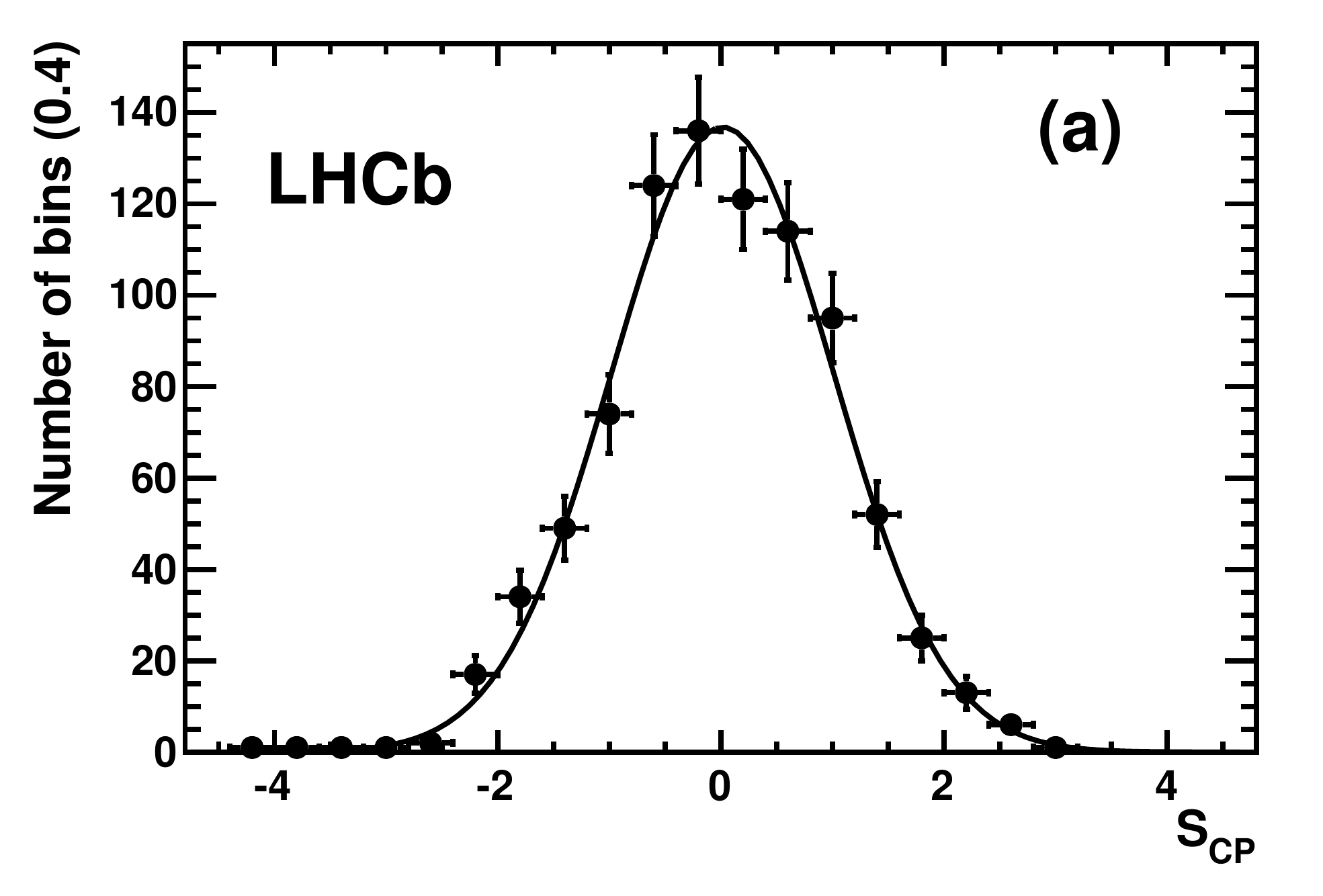}
  \includegraphics[width=0.45\textwidth]{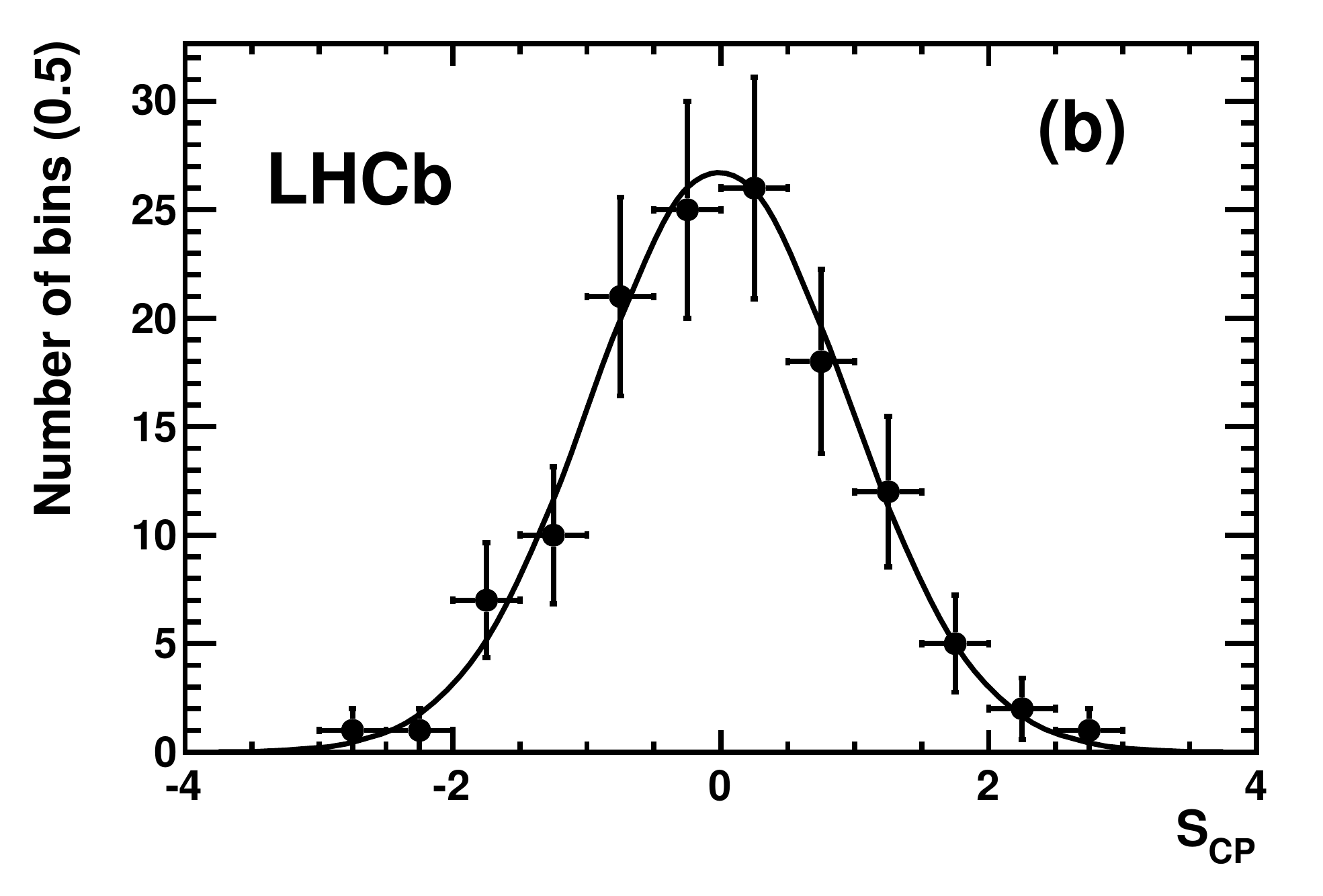}
\end{center}
\caption{(a) Distribution of $\mathcal{S}_{CP}$ values from $\Dp \to \Km\pip\pip$ from a test with 900 
  uniform bins. The mean of the fitted Gaussian distribution is
  $0.015\pm0.034$ and the width is $0.996\pm0.023$. (b)
  Distribution of $\mathcal{S}_{CP}$ values from  $\Dsp \to \Km\Kp\pip$ with
  129 bins. The fitted mean is $-0.011\pm0.084$ and the
  width is $0.958\pm0.060$.
}
\label{fig:control:miranda}
\end{figure*}

\begin{figure*}[tbp]
\begin{center}
\includegraphics[width=0.45\textwidth]{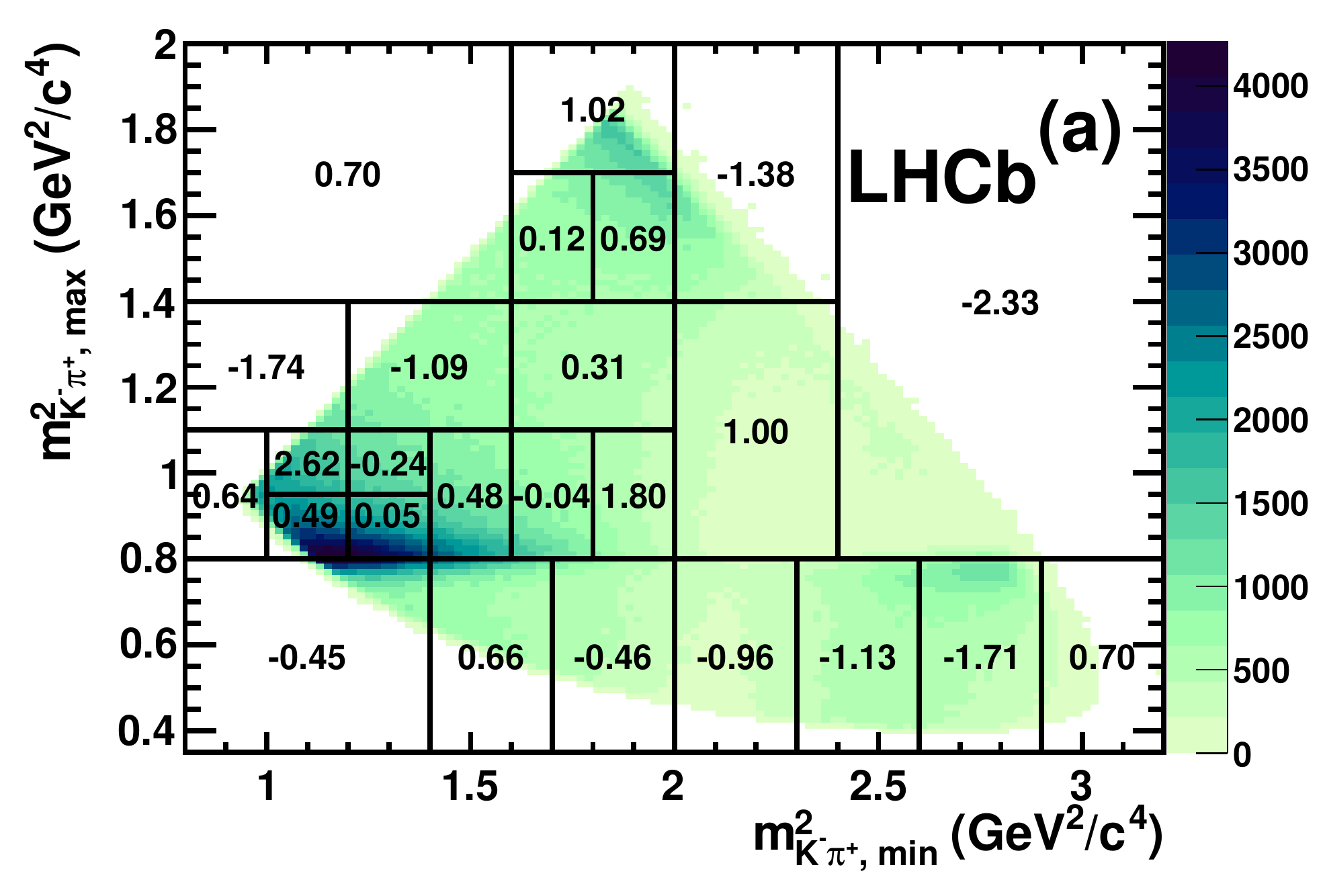}
\includegraphics[width=0.45\textwidth]{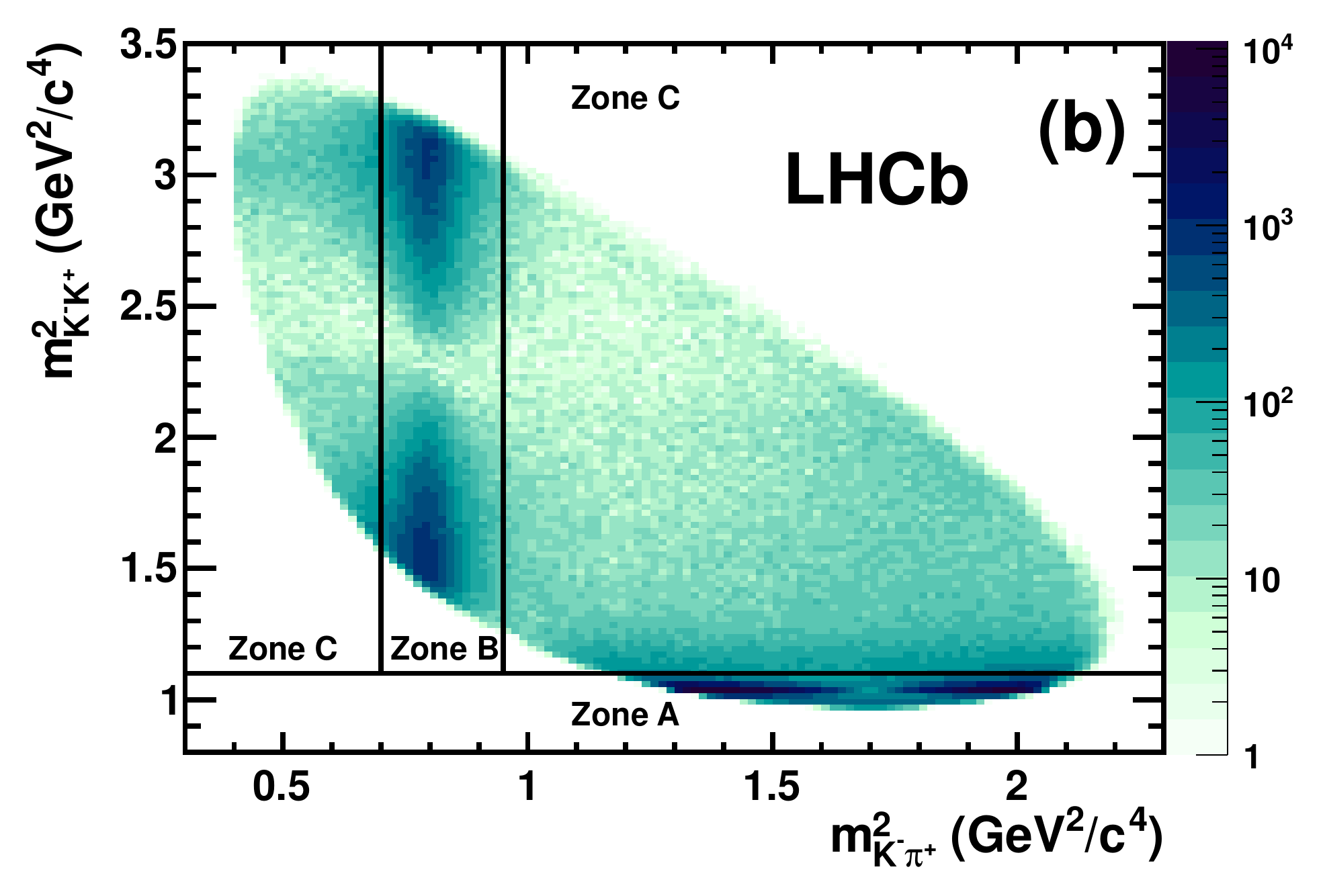}
\end{center}
\caption{
  Dalitz plots of (a)  $\Dp \to \Km\pip\pip$, showing the
  25-bin adaptive scheme with the $\mathcal{S}_{CP}$ values,  and (b)
  $\Dsp \to \Km\Kp\pip$, showing the three regions referred to in the
  text. The higher and lower $\Km\pip$ invariant mass combinations
  are plotted in (a) as there are identical pions in the final state.
  }
  \label{fig:control:dalitz}
\end{figure*}

The control modes and their mass sidebands defined in Fig.~\ref{fig:data:massfit} are tested for asymmetries using the method described in the previous
section. Adaptive and uniform binning schemes are defined for  $\Dp
\to \Km\pip\pip$ and $\Dsp \to \Km\Kp\pip$. They are applied to
samples 1--3 and each magnet polarity separately. In the final
results, the asymmetries measured in data taken with positive and negative magnet
polarity are combined in order to cancel left-right detector asymmetries. The precise number of bins chosen is
arbitrary, but care is taken to use a wide range of tests with
binnings that reflect the size of the dataset for the decay mode under study.

 For  $\Dp \to \Km\pip\pip$, five different sets of bins
 in each scheme are used. A very low $p$-value would indicate a local asymmetry. One test with 25 adaptive bins in 
one of the subsamples (with negative magnet polarity) has a $p$-value
of 0.1\%, but when combined with the positive polarity sample the
$p$-value increases to 1.7\%. All other tests yield $p$-values
ranging from 1--98\%. Some example results are given in Table~\ref{table:control:kpp-uniform}.  A typical distribution of the
$\mathcal{S}_{CP}$ values with a Gaussian fit is shown in
Fig.~\ref{fig:control:miranda}(a) for a test with 900 uniform bins. The fitted values of the mean and
width are consistent with one and zero respectively, suggesting that the differences between the $D^+$ and the $D^-$ 
Dalitz plots are driven only by statistical fluctuations.

\begin{table}[tbp]
 
    \caption{Results ($p$-values, in \%) from tests with the $\Dp \to \Km\pip\pip$ 
    control channel using the uniform and adaptive binning schemes. The values correspond to
    tests performed on the whole dataset in the mass windows defined
    in Sect.~\ref{sec:data}.     
    \label{table:control:kpp-uniform}}

 \begin{center}
\begin{tabular}{lcccccc}   
\hline\hline
     & 1300 bins   &  900 bins  &  400 bins  &   100 bins  & 25 bins  \\ \hline 
Uniform &     73.8    &   17.7     &    72.6    &     54.6    &  1.7 \\ 
Adaptive&     81.7    &   57.4     &    65.8    &    30.0     &  11.8\\
\hline \hline
\end{tabular}
  \end{center}
  
\end{table}

For the $\Dsp \to \Km\Kp\pip$ mode a different procedure is
followed due to the 
smaller sample size and to the high density of events along
the $\phi$ and the $\Kbar^*(892)^{0}$ bands. The Dalitz plot is divided into three zones, as shown in Fig.~\ref{fig:control:dalitz}.
 Each zone is further divided into 300, 
100 and 30 bins of same size. The results are given in Table~\ref{table:control:kkp1}. 
In addition, a test is performed on the whole Dalitz plot using
129 bins chosen by the adaptive algorithm, and a version of
the 25-bin scheme outlined in Sect. \ref{sec:toys} scaled by the ratio
of the available phase space in the two modes. These tests
yield $p$-values of 71.5\% and 34.3\% respectively.

\begin{table}[tbp]
    \caption{Results ($p$-values, in \%)  from tests with the $\Dsp \to \Km\Kp\pip$ 
    control channel using the uniform binning scheme. The values correspond to
    tests performed separately on Zones A-C, with samples 1-3 and both
    magnet polarities combined.     
    \label{table:control:kkp1}}

\begin{center}
\begin{tabular}{lcccc}   
\hline\hline

 bins  & Zone A   & Zone B  & Zone C     \\ \hline
300    &     20.1        &   25.3 	  &    14.5	\\
100    &     41.7        &   84.6 	  &    89.5	\\ 
$\phantom{0}$30     &     66.0        &   62.5 	  &    24.6	\\
\hline\hline
\end{tabular}
  \end{center}
  
\end{table}

Other possible
sources of local charge asymmetry in the signal region 
are the charm contamination of the background, and asymmetries from
CPV in misreconstructed $B$ decays. In order to 
investigate the first possibility, similar tests are carried out in the mass
sidebands of the $D^+_{(s)} \to \Km\Kp\pip$ signal
(illustrated in Fig.~\ref{fig:data:massfit}).
There is no evidence for asymmetries in the background.

From a simulation of the decay $\Dp \to \Km\pip\pip$
the level of secondary charm ($B \to DX$) in our selected sample is found to be 4.5\%. The main discriminating variable to distinguish between prompt and secondary
charm is the impact parameter (IP) of the $D$ with respect to 
the primary vertex. Given the long $B$ lifetime, the IP distribution
of secondary charm candidates is shifted towards larger values compared to that of prompt 
$\Dp$ mesons. 

The effect of secondary charm is investigated by dividing the data set 
according to the value of the candidate IP significance ($\chi^{2}_{IP}$).
The subsample with events having larger $\chi^{2}_{IP}$ are likely to be 
richer in secondary charm. The results are shown in Table~\ref{ipchi2}. No anomalous effects are seen in the high $\chi^{2}_{IP}$ sample, so contamination from secondary charm with CPV does not affect
our results for studies with our current level of sensitivity.

\begin{table}[tbp]
  \caption{Results ($p$-values, in \%)  from tests with the $\Dp \to \Km\pip\pip$ 
   and  $\Dsp \to \Km\Kp\pip$ samples divided according to the impact parameter
   with respect to the primary vertex. The tests are performed
   using the adaptive binning scheme with 25 bins. 
  }
  \begin{center}
    \begin{tabular}{ccc}
\hline\hline
                   &$\chi^{2}_{IP}<6$ & $\chi^{2}_{IP}>6$ \\ \hline
   
$\Dp \to \Km\pip\pip$  &    8.5  &   88.9   \\
$\Dsp \to \Km\Kp\pip$    &   52.0  &   30.6   \\
\hline\hline

    \end{tabular}
  \end{center}
  \label{ipchi2}
\end{table}

The analysis on the two control modes and on
the sidebands in the final states $\Km\Kp\pip$ and $\Km\pip\pip$ gives results from all tests that are fully consistent with no asymmetry.  
Therefore, any asymmetry observed in $\Dp \to \Km\Kp\pip$ is
likely to be a real physics effect.

\begin{table*}[tbp]
\caption{Fitted means and widths, $\chi^2/{\rm ndf}$ and $p$-values
  for consistency with no CPV for the $\Dp \to \Km\Kp\pip$ decay
  mode with four different binnings.
\label{results:pvalues}}
\begin{center}
\begin{tabular}{ccccc}
\hline\hline
  Binning    & Fitted mean &  Fitted width &   $\chi^2/{\rm ndf}$  &   $p$-value (\%)  \\ \hline
 Adaptive I  & $\phantom{-}0.01\pm0.23$ & $1.13\pm0.16$  &      32.0/24     & 12.7 \\
Adaptive II &  $-0.024\pm0.010$& $1.078\pm0.074$&   123.4/105    & 10.6    \\ 
 Uniform I   & $-0.043\pm0.073$ & $0.929\pm0.051$  &  191.3/198    & 82.1   \\   
 Uniform II  & $-0.039\pm0.045$ & $1.011\pm0.034$ &     519.5/529    & 60.5  \\     
\hline\hline
\end{tabular}
\end{center}

\end{table*}

\begin{figure*}[tbp]
\begin{center}

\includegraphics*[width=0.45\textwidth]{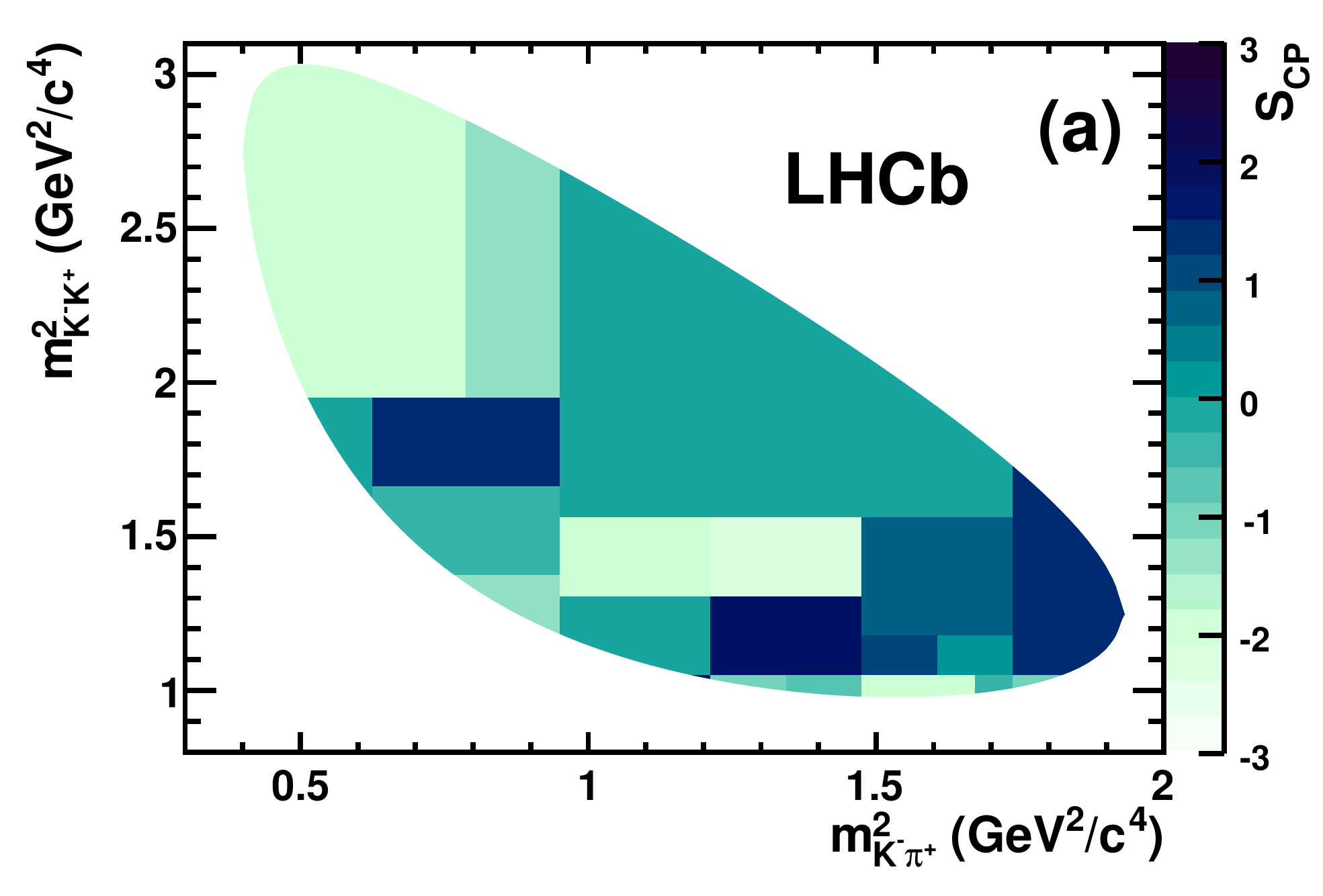}\includegraphics*[width=0.45\textwidth]{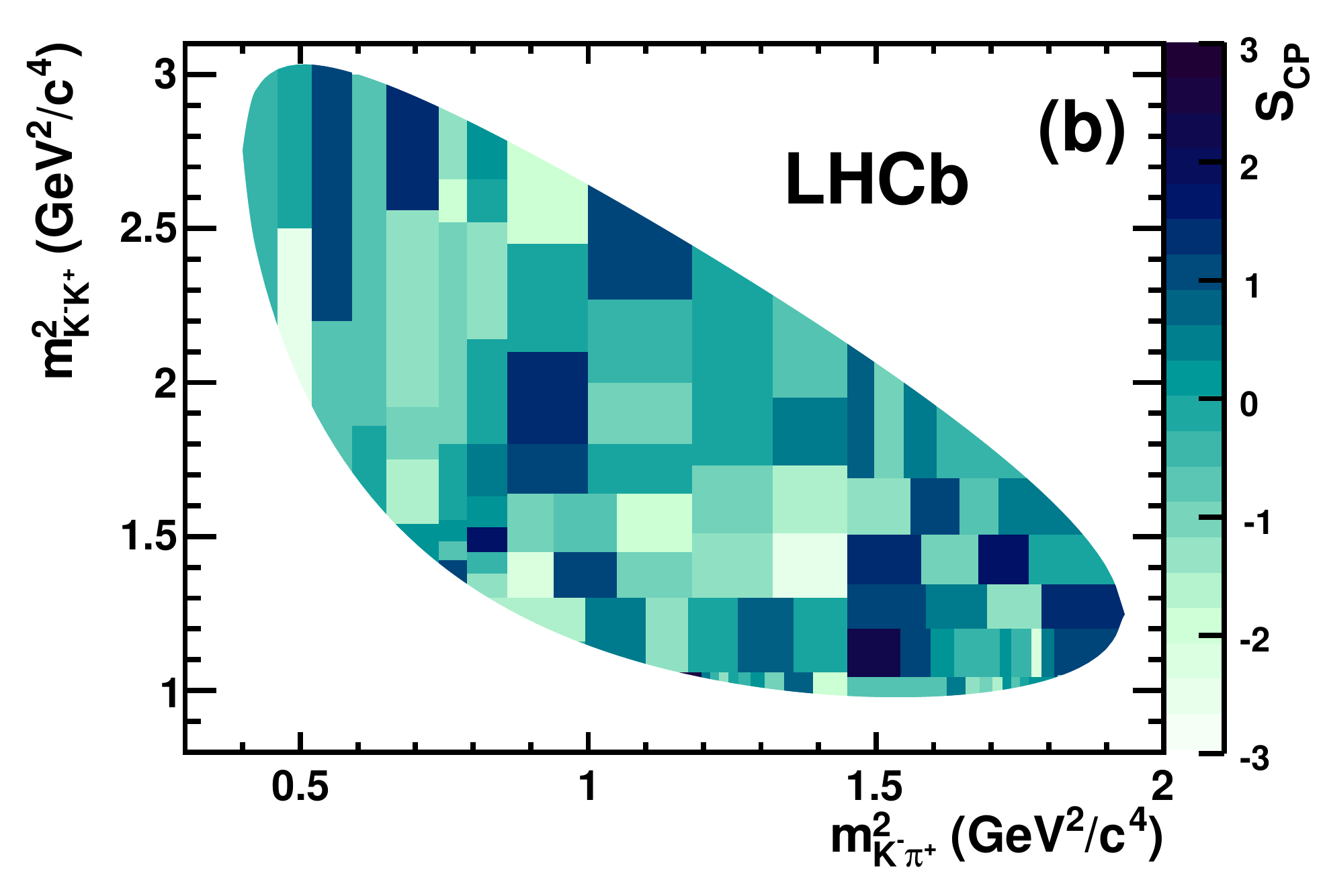}

\includegraphics*[width=0.45\textwidth]{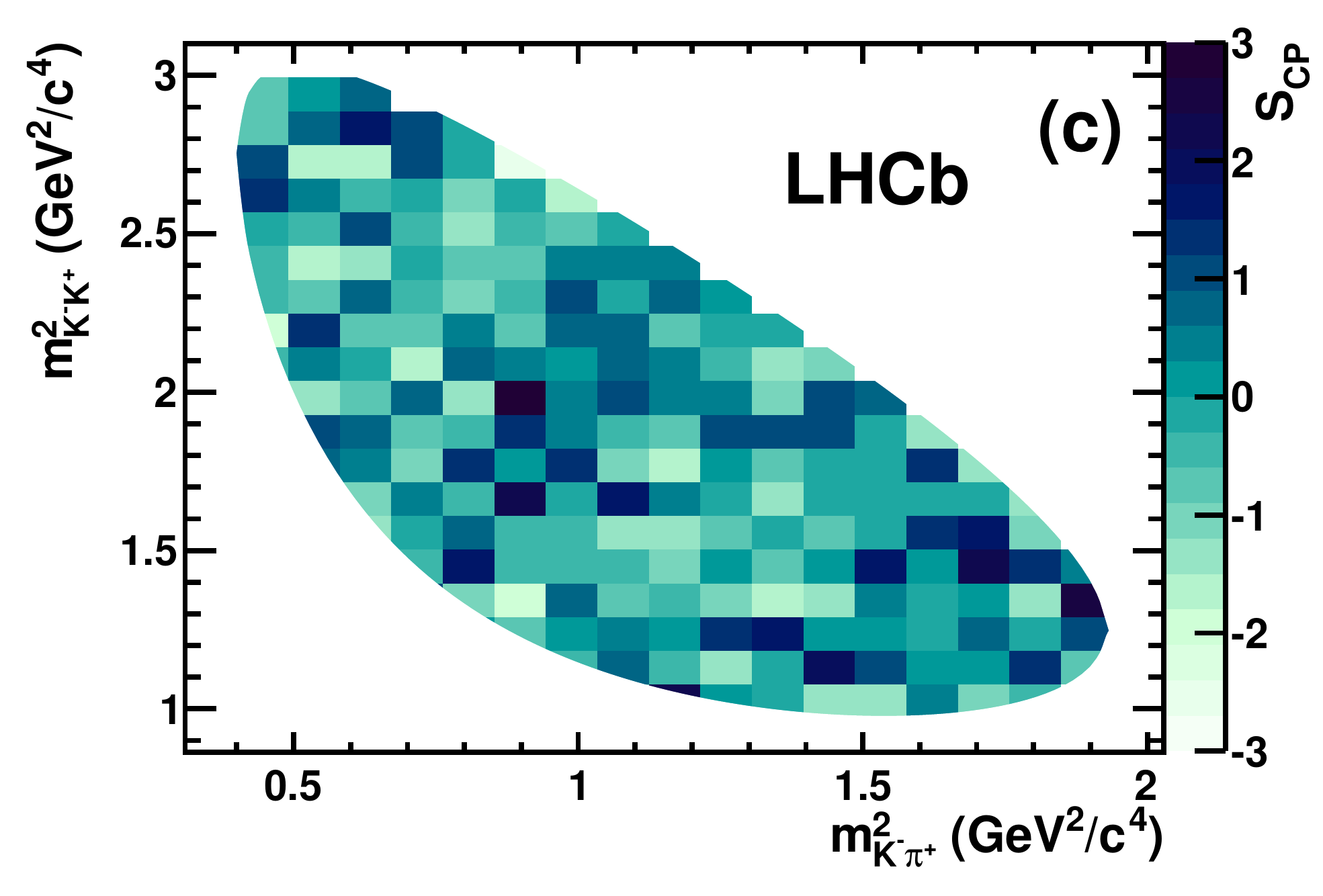}\includegraphics*[width=0.45\textwidth]{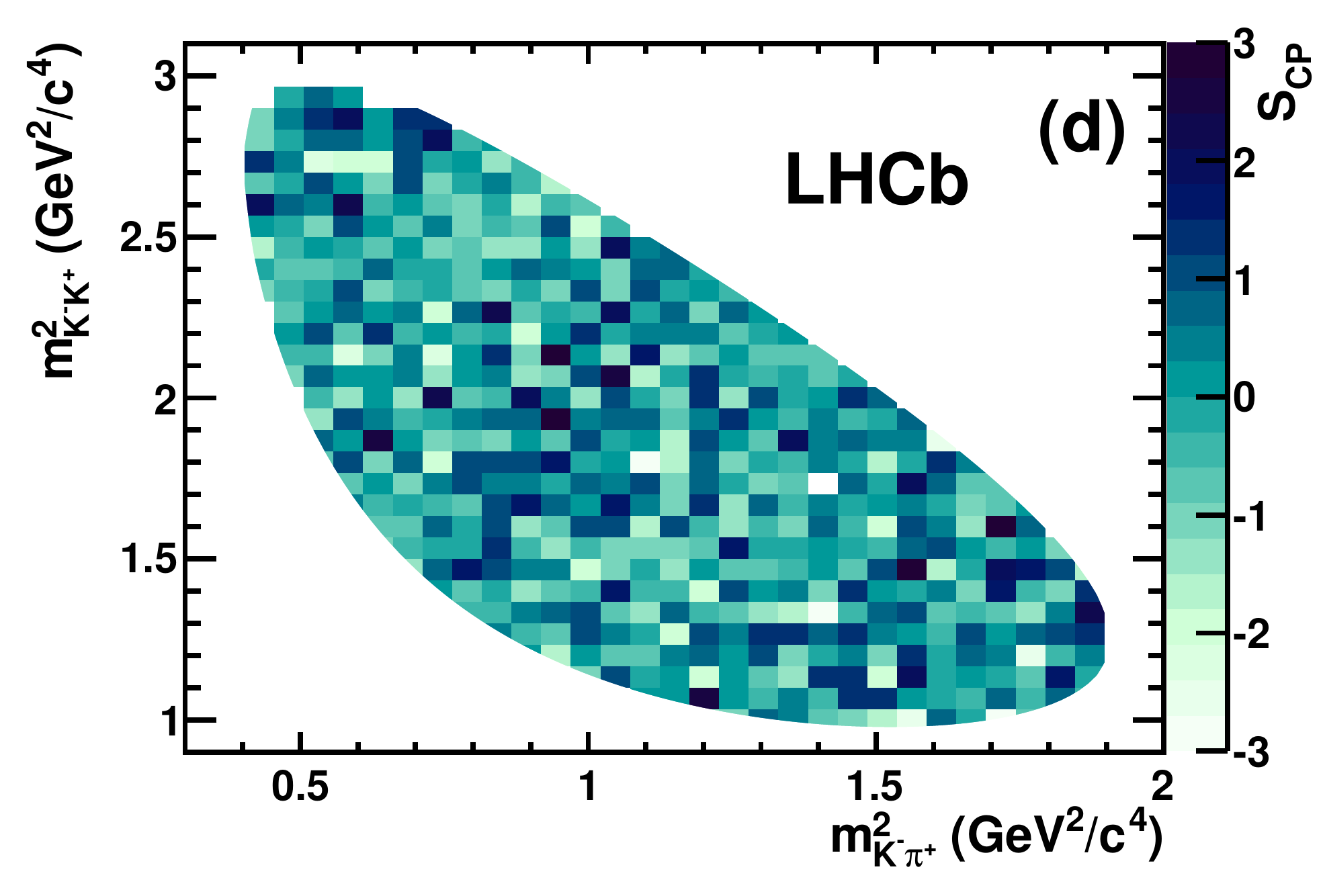}
\end{center}

\caption{Distribution of ${\mathcal S}^i_{\it CP}$ in
  the Dalitz plot for (a) ``Adaptive I",  (b) ``Adaptive II'', (c)
  ``Uniform I'' and (d) ``Uniform II". In (c) and (d) bins at the edges are
 not shown if the number of entries is not above a threshold of 50
  (see Sect.~\ref{sec:toys}).}
\label{results:scp-dalitz}
\end{figure*}

\begin{figure*}[tbp]
\begin{center}

\includegraphics*[width=0.45\textwidth]{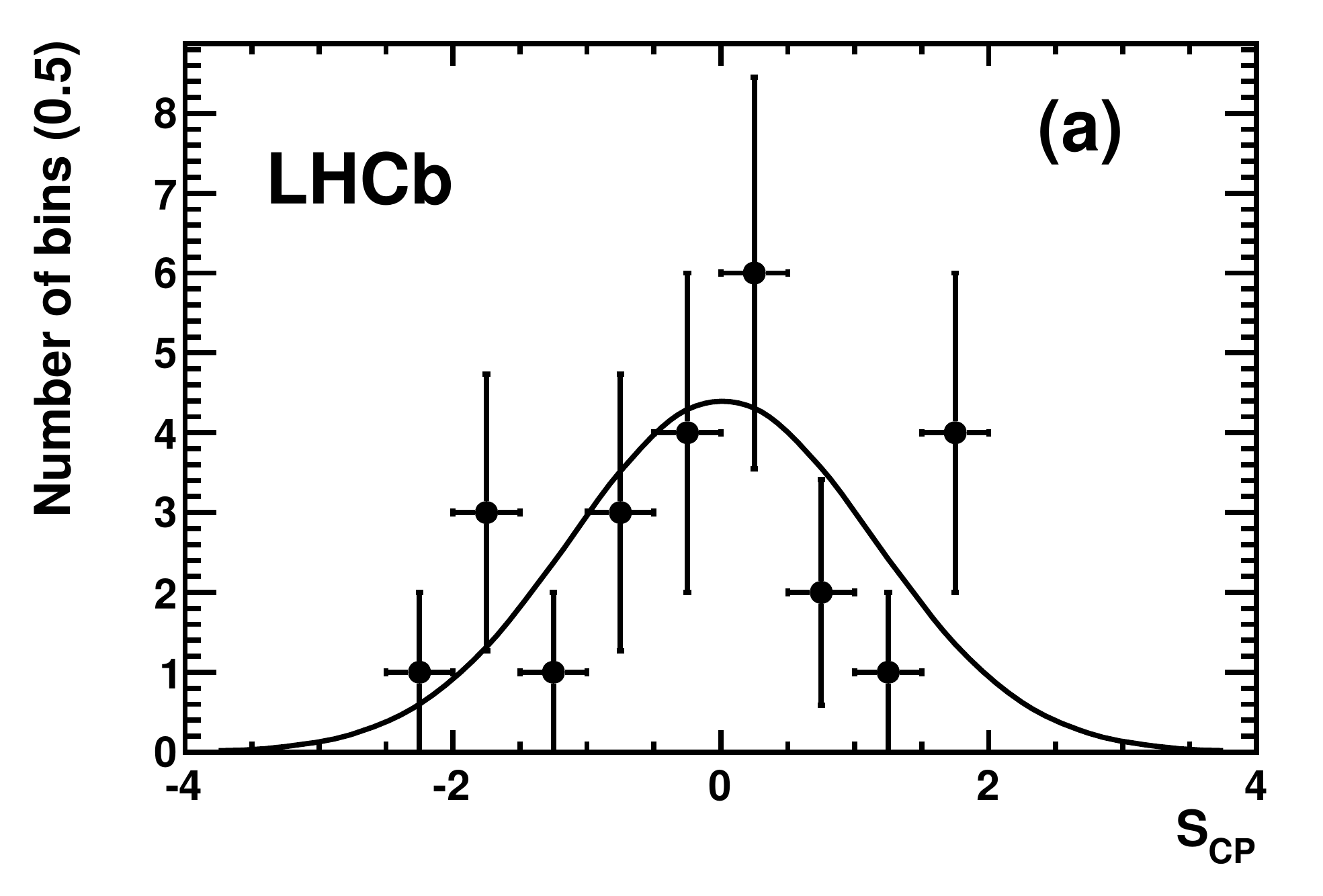}
\includegraphics*[width=0.45\textwidth]{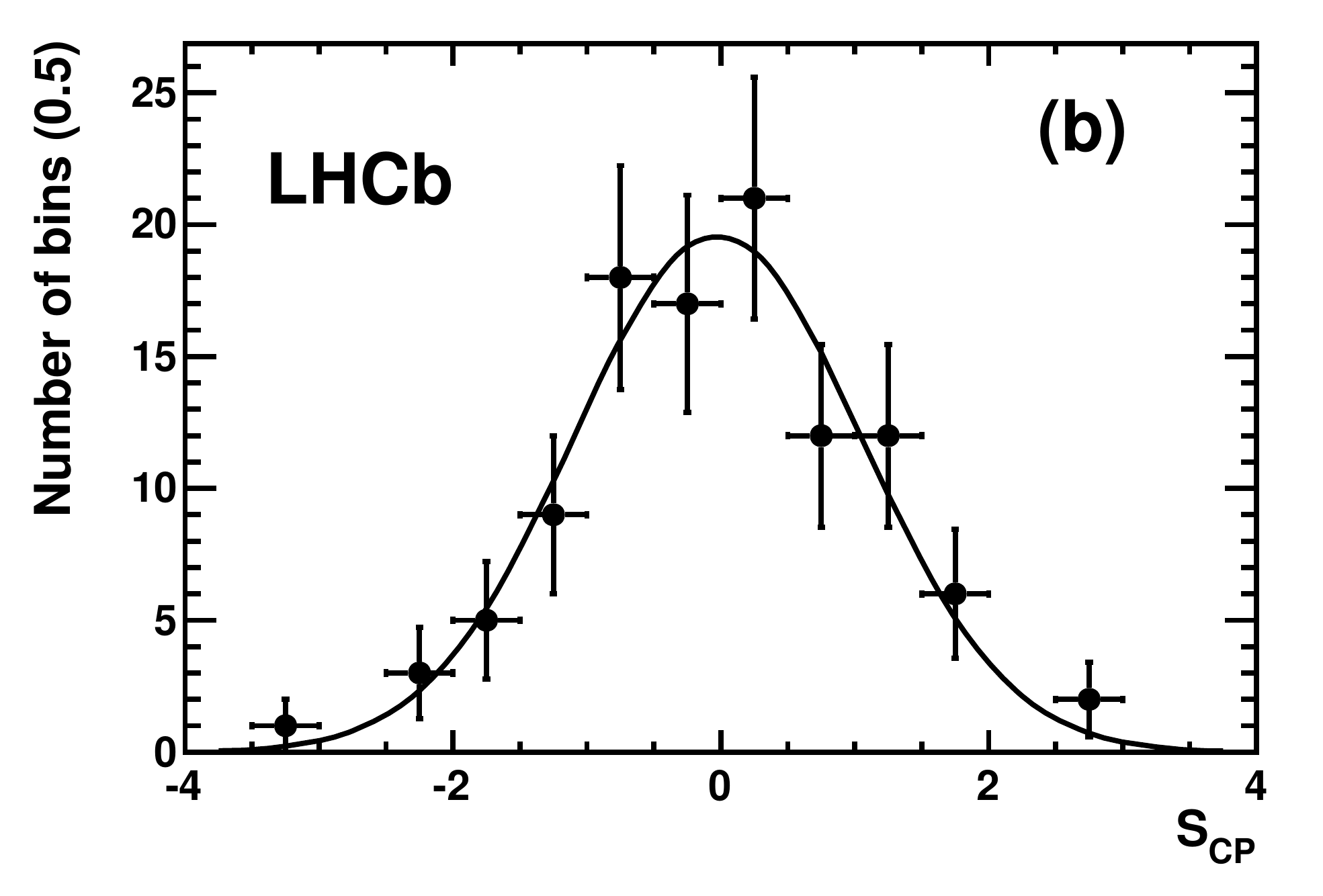}
\includegraphics*[width=0.45\textwidth]{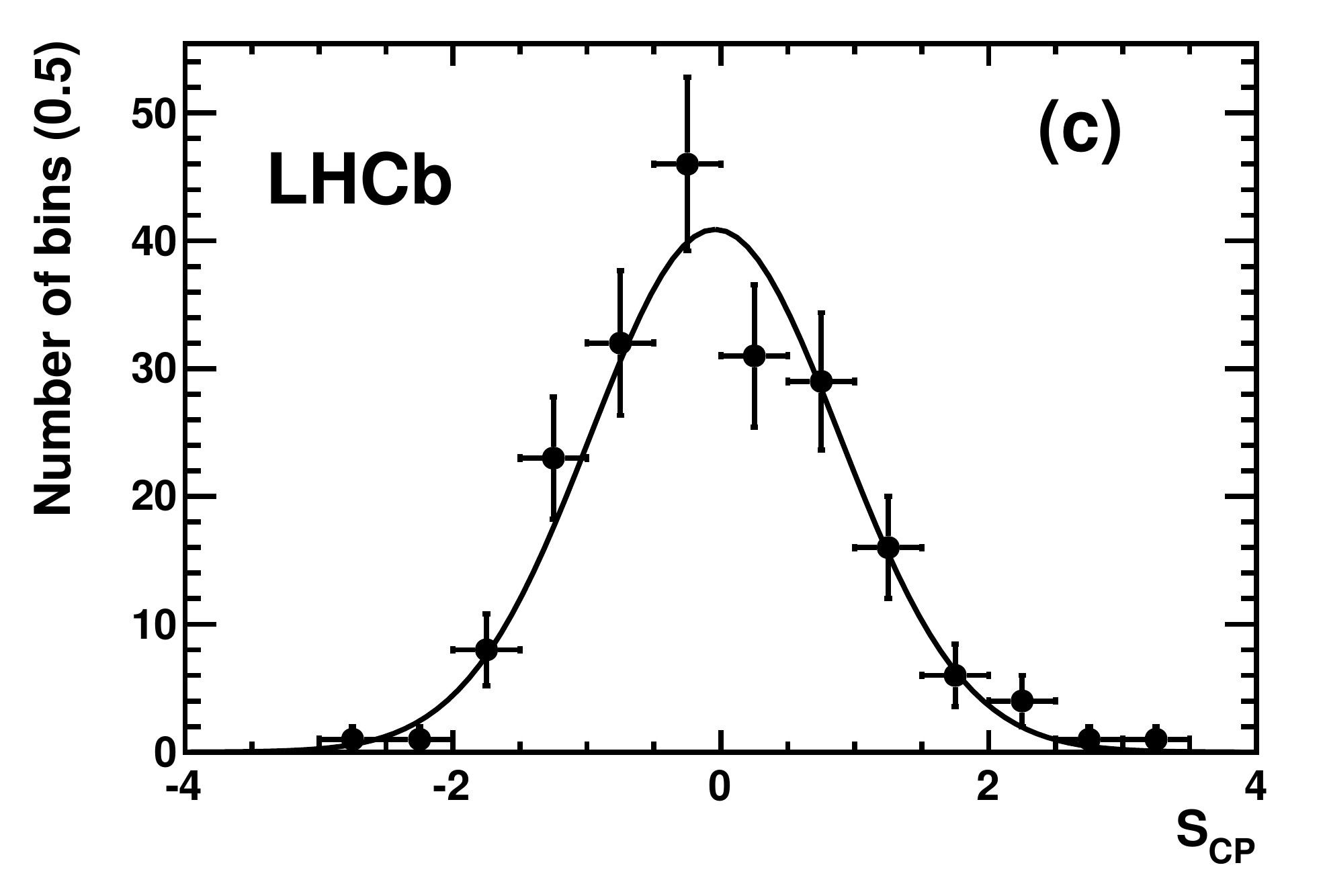}
\includegraphics*[width=0.45\textwidth]{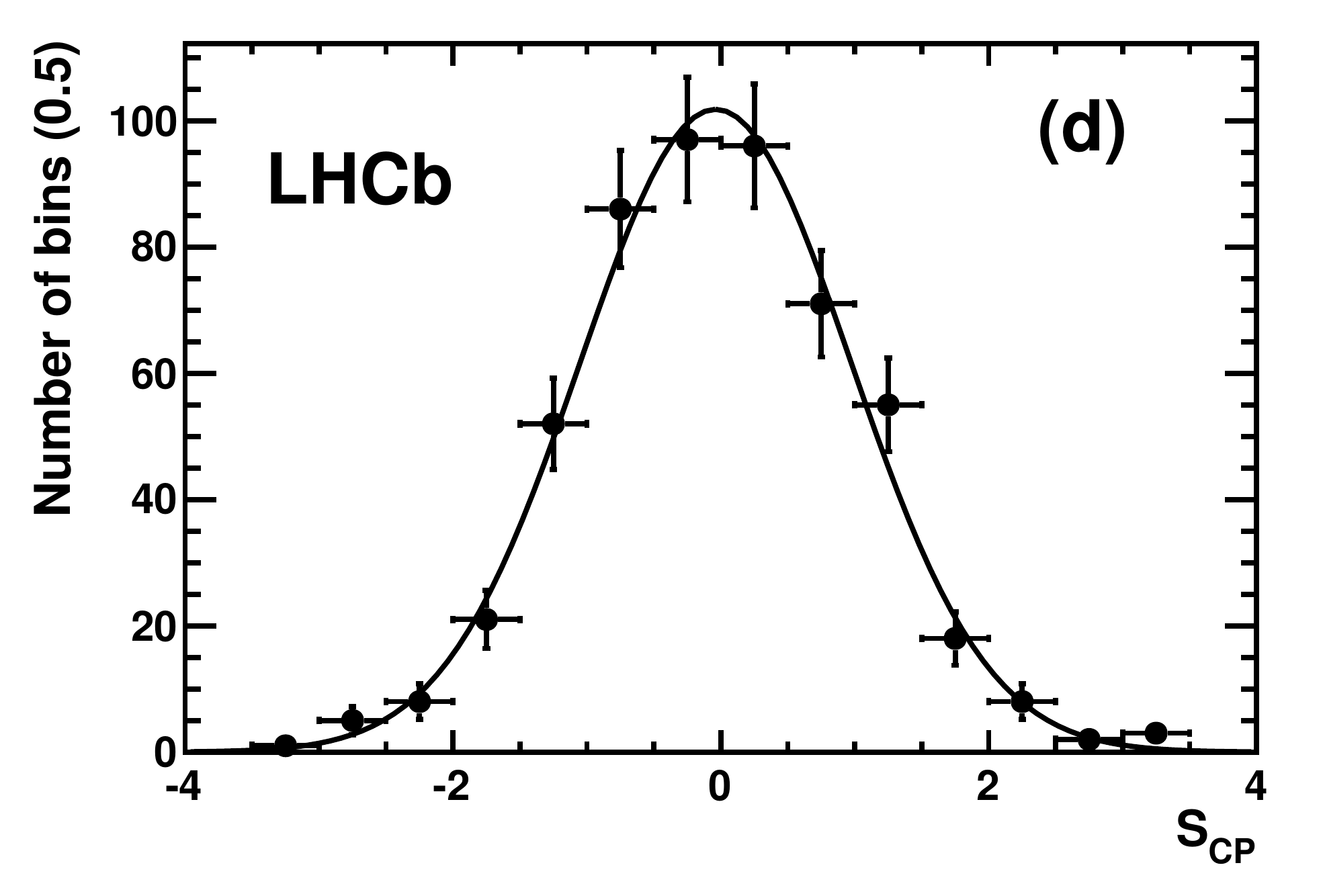}
\end{center}
\caption{Distribution of ${\mathcal S}^i_{\it CP}$  fitted to Gaussian
  functions,
  for  (a) ``Adaptive I", (b) ``Adaptive II'', (c) ``Uniform I'' and
  (d) ``Uniform II". The fit results are given in Table~\ref{results:pvalues}.}
\label{results:miranda530_gaus}
\end{figure*}

\section{Results}
\label{sec:results}

The  signal sample with which we search for \CP violation consists of
403,894 candidates selected within the $\Km\Kp\pip$ mass window from 1856.7 to 1882.1~MeV$/c^2$, as described in Sect.~\ref{sec:data}. There are 200,336 and
203,558 $\Dp$ and  $\Dm$ candidates 
respectively. This implies a normalization factor $\alpha = N_{\rm tot}(\Dp)/N_{\rm tot}(\Dm) = 0.984\pm 0.003$, to be used in  Eq.~\ref{intro:signif}.

The  strategy for looking for signs of localized CPV
is discussed in the previous sections. In the absence of local asymmetries in the control channels $\Dp \to \Km\pip\pip$ and $\Dsp \to \Km\Kp\pip$ and
in the sidebands of the $\Km\Kp\pip$ mass spectrum, we investigate the signal sample under different binning choices. 

 First, the adaptive binning is used with 25 and 106 bins in the
 Dalitz plot as illustrated in Fig.~\ref{fig:toydist1}.
Then CPV is investigated with uniform binnings, using 200 and 530 bins of equal size. For each of these binning choices, the significance
${\cal S}^i_{\it CP}$ of the difference in $\Dp$ and $\Dm$ population
is computed for each bin $i$, as defined in Eq.~\ref{intro:signif}. The $\chi^2/{\rm ndf} = \sum_i ({\cal S}^i_{\it CP})^{2}/{\rm ndf}$
is calculated and the $p$-value is obtained. The distributions of
${\mathcal S}^i_{\it CP}$ are fitted to Gaussian functions. 

The $p$-values are shown in Table~\ref{results:pvalues}.  The Dalitz plot distributions of ${\mathcal S}^i_{\it CP}$ are shown 
in Fig.~\ref{results:scp-dalitz}. In
Fig.~\ref{results:miranda530_gaus} the distributions of ${\mathcal
  S}^i_{\it CP}$ and the corresponding Gaussian fits for the different
binnings are shown. The $p$-values obtained
indicate no evidence for CPV. This is corroborated by the good fits of the ${\mathcal S}^i_{\it CP}$
distributions to Gaussians, with means and widths consistent with 0 and 1, respectively. 

As further checks, many other binnings are tested. The number of bins
in the adaptive and uniform binning schemes is varied from 28 to 106
and from 21 to 530 respectively. The samples are separated according
to the magnet polarity and the same studies are repeated. In all cases
the $p$-values are consistent with no CPV, with values ranging from
4\% to 99\%.  We conclude that there is no evidence for CPV in our
data sample of $\Dp \to \Km\Kp\pip$.
\section{Conclusion}
\label{sec:conclusion}
Due to the rich structure of their Dalitz plots, three body charm
decays are sensitive to \CP violating phases within and beyond the
Standard Model. Here, a model-independent search for direct \CP violation is performed
in the Cabibbo suppressed decay $\Dp \to \Km\Kp\pip$
with 35~pb$^{-1}$ of data collected by the \lhcb experiment, and no
evidence for CPV is found. Several binnings
are used to compare normalised  \Dp and \Dm Dalitz
plot distributions. This technique is validated with
large numbers of simulated pseudo-experiments and with Cabibbo favoured
control channels from the data: no false positive signals are seen. To our knowledge this is the first time a
search for CPV is performed using adaptive bins which reflect
the structure of the Dalitz plot.

Monte Carlo simulations illustrate that large localised asymmetries 
can occur without causing detectable differences in integrated
decay rates. The technique used here is shown to be sensitive to such
asymmetries. Assuming the decay model, efficiency parameterisation and
background model described in Sect.~\ref{sec:toys} we would be 90\%
confident of seeing a \CP violating difference of either $5^{\circ}$ in the phase of
the $\phi\pip$ or 11\%
 in the magnitude of the $\kappa(800)K^{+}$ with
$3\sigma$ significance. Since we find no evidence of CPV, effects of
this size are unlikely to exist.

\section{Acknowledgments}

We express our gratitude to our colleagues in the CERN accelerator
departments for the excellent performance of the LHC. We thank the
technical and administrative staff at CERN and at the LHCb institutes,
and acknowledge support from the National Agencies: CAPES, CNPq,
FAPERJ and FINEP (Brazil); CERN; NSFC (China); CNRS/IN2P3 (France);
BMBF, DFG, HGF and MPG (Germany); SFI (Ireland); INFN (Italy); FOM and
NWO (Netherlands); SCSR (Poland); ANCS (Romania); MinES of Russia and
Rosatom (Russia); MICINN, XuntaGal and GENCAT (Spain); SNSF and SER
(Switzerland); NAS Ukraine (Ukraine); STFC (United Kingdom); NSF
(USA). We also acknowledge the support received from the ERC under FP7
and the Region Auvergne.

\bibliographystyle{LHCb}
\bibliography{refs}

\ifx\mcitethebibliography\mciteundefinedmacro
\PackageError{LHCb.bst}{mciteplus.sty has not been loaded}
{This bibstyle requires the use of the mciteplus package.}\fi
\providecommand{\href}[2]{#2}
\begin{mcitethebibliography}{10}
\mciteSetBstSublistMode{n}
\mciteSetBstMaxWidthForm{subitem}{\alph{mcitesubitemcount})}
\mciteSetBstSublistLabelBeginEnd{\mcitemaxwidthsubitemform\space}
{\relax}{\relax}
\bibitem{Cabibbo:1963yz}
N.~Cabibbo, \ifthenelse{\boolean{articletitles}}{{\it {Unitary Symmetry and
  Leptonic Decays}}, }{} \href{http://dx.doi.org/10.1103/PhysRevLett.10.531}{
  {\em Phys. Rev. Lett.} {\bf 10} (1963) 531--533}\relax
\mciteBstWouldAddEndPuncttrue
\mciteSetBstMidEndSepPunct{\mcitedefaultmidpunct}
{\mcitedefaultendpunct}{\mcitedefaultseppunct}\relax
\EndOfBibitem
\bibitem{Kobayashi:1973fv}
M.~Kobayashi and T.~Maskawa, \ifthenelse{\boolean{articletitles}}{{\it {CP
  Violation in the Renormalizable Theory of Weak Interaction}}, }{}
  \href{http://dx.doi.org/10.1143/PTP.49.652}{ {\em Prog. Theor. Phys.} {\bf
  49} (1973) 652--657}\relax
\mciteBstWouldAddEndPuncttrue
\mciteSetBstMidEndSepPunct{\mcitedefaultmidpunct}
{\mcitedefaultendpunct}{\mcitedefaultseppunct}\relax
\EndOfBibitem
\bibitem{Bianco:2003vb}
S.~Bianco, F.~L. Fabbri, D.~Benson, and I.~Bigi,
  \ifthenelse{\boolean{articletitles}}{{\it {{A Cicerone for the physics of
  charm}}}, }{} {\em Riv. Nuovo Cim.} {\bf 26N7} (2003) 1--200,
  [\href{http://xxx.lanl.gov/abs/hep-ex/0309021}{{arXiv:hep-ex/0309021}}]\relax
\mciteBstWouldAddEndPuncttrue
\mciteSetBstMidEndSepPunct{\mcitedefaultmidpunct}
{\mcitedefaultendpunct}{\mcitedefaultseppunct}\relax
\EndOfBibitem
\bibitem{Artuso:2008vf}
M.~Artuso, B.~Meadows, and A.~A. Petrov,
  \ifthenelse{\boolean{articletitles}}{{\it {Charm Meson Decays}}, }{}
  \href{http://dx.doi.org/10.1146/annurev.nucl.58.110707.171131}{ {\em Ann.
  Rev. Nucl. Part. Sci.} {\bf 58} (2008) 249--291},
  [\href{http://xxx.lanl.gov/abs/0802.2934}{{arXiv:0802.2934}}]\relax
\mciteBstWouldAddEndPuncttrue
\mciteSetBstMidEndSepPunct{\mcitedefaultmidpunct}
{\mcitedefaultendpunct}{\mcitedefaultseppunct}\relax
\EndOfBibitem
\bibitem{Grossman:2006jg}
Y.~Grossman, A.~L. Kagan, and Y.~Nir, \ifthenelse{\boolean{articletitles}}{{\it
  {New physics and CP violation in singly Cabibbo suppressed D decays}}, }{}
  \href{http://dx.doi.org/10.1103/PhysRevD.75.036008}{ {\em Phys. Rev.} {\bf
  D75} (2007) 036008},
  [\href{http://xxx.lanl.gov/abs/hep-ph/0609178}{{arXiv:hep-ph/0609178}}]\relax
\mciteBstWouldAddEndPuncttrue
\mciteSetBstMidEndSepPunct{\mcitedefaultmidpunct}
{\mcitedefaultendpunct}{\mcitedefaultseppunct}\relax
\EndOfBibitem
\bibitem{Aubert:2005gj}
BABAR Collaboration, B.~Aubert et~al.,
  \ifthenelse{\boolean{articletitles}}{{\it {A search for CP violation and a
  measurement of the relative branching fraction in $D^+ \to K^- K^+ \pi^+$
  decays}}, }{} \href{http://dx.doi.org/10.1103/PhysRevD.71.091101}{ {\em Phys.
  Rev.} {\bf D71} (2005) 091101},
  [\href{http://xxx.lanl.gov/abs/hep-ex/0501075}{{arXiv:hep-ex/0501075}}]\relax
\mciteBstWouldAddEndPuncttrue
\mciteSetBstMidEndSepPunct{\mcitedefaultmidpunct}
{\mcitedefaultendpunct}{\mcitedefaultseppunct}\relax
\EndOfBibitem
\bibitem{:2008zi}
CLEO Collaboration, P.~Rubin et~al., \ifthenelse{\boolean{articletitles}}{{\it
  {Search for CP Violation in the Dalitz-Plot Analysis of $D^+ \to K^- K^+
  \pi^+$}}, }{} \href{http://dx.doi.org/10.1103/PhysRevD.78.072003}{ {\em Phys.
  Rev.} {\bf D78} (2008) 072003},
  [\href{http://xxx.lanl.gov/abs/0807.4545}{{arXiv:0807.4545}}]\relax
\mciteBstWouldAddEndPuncttrue
\mciteSetBstMidEndSepPunct{\mcitedefaultmidpunct}
{\mcitedefaultendpunct}{\mcitedefaultseppunct}\relax
\EndOfBibitem
\bibitem{:2011en}
Belle Collaboration, M.~Stari\v{c} et~al., \ifthenelse{\boolean{articletitles}}{{\it {Search
  for CP Violation in $D$ Meson Decays to $\phi \pi^{+}$}}, }{}
  \href{http://xxx.lanl.gov/abs/1110.0694}{{arXiv:1110.0694}}\relax
\mciteBstWouldAddEndPuncttrue
\mciteSetBstMidEndSepPunct{\mcitedefaultmidpunct}
{\mcitedefaultendpunct}{\mcitedefaultseppunct}\relax
\EndOfBibitem
\bibitem{Bediaga:2009tr}
I.~Bediaga et~al., \ifthenelse{\boolean{articletitles}}{{\it {On a CP
  anisotropy measurement in the Dalitz plot}}, }{}
  \href{http://dx.doi.org/10.1103/PhysRevD.80.096006}{ {\em Phys. Rev.} {\bf
  D80} (2009) 096006},
  [\href{http://xxx.lanl.gov/abs/0905.4233}{{arXiv:0905.4233}}]\relax
\mciteBstWouldAddEndPuncttrue
\mciteSetBstMidEndSepPunct{\mcitedefaultmidpunct}
{\mcitedefaultendpunct}{\mcitedefaultseppunct}\relax
\EndOfBibitem
\bibitem{Aubert:2008yd}
BABAR Collaboration, B.~Aubert et~al.,
  \ifthenelse{\boolean{articletitles}}{{\it {Search for CP Violation in Neutral
  D Meson Cabibbo-suppressed Three-body Decays}}, }{}
  \href{http://dx.doi.org/10.1103/PhysRevD.78.051102}{ {\em Phys. Rev.} {\bf
  D78} (2008) 051102},
  [\href{http://xxx.lanl.gov/abs/0802.4035}{{arXiv:0802.4035}}]\relax
\mciteBstWouldAddEndPuncttrue
\mciteSetBstMidEndSepPunct{\mcitedefaultmidpunct}
{\mcitedefaultendpunct}{\mcitedefaultseppunct}\relax
\EndOfBibitem
\bibitem{lyons1989statistics}
L.~Lyons, {\em Statistics for nuclear and particle physicists}.
\newblock Cambridge University Press, 1989\relax
\mciteBstWouldAddEndPuncttrue
\mciteSetBstMidEndSepPunct{\mcitedefaultmidpunct}
{\mcitedefaultendpunct}{\mcitedefaultseppunct}\relax
\EndOfBibitem
\bibitem{Alves:2008zz}
LHCb Collaboration, A.~Alves et~al., \ifthenelse{\boolean{articletitles}}{{\it
  {The LHCb Detector at the LHC}}, }{}
  \href{http://dx.doi.org/10.1088/1748-0221/3/08/S08005}{ {\em JINST} {\bf 3}
  (2008) S08005}\relax
\mciteBstWouldAddEndPuncttrue
\mciteSetBstMidEndSepPunct{\mcitedefaultmidpunct}
{\mcitedefaultendpunct}{\mcitedefaultseppunct}\relax
\EndOfBibitem
\bibitem{Agostinelli:2002hh}
GEANT4, S.~Agostinelli et~al., \ifthenelse{\boolean{articletitles}}{{\it
  {GEANT4: A Simulation toolkit}}, }{}
  \href{http://dx.doi.org/10.1016/S0168-9002(03)01368-8}{ {\em Nucl. Instrum.
  Meth.} {\bf A506} (2003) 250--303}\relax
\mciteBstWouldAddEndPuncttrue
\mciteSetBstMidEndSepPunct{\mcitedefaultmidpunct}
{\mcitedefaultendpunct}{\mcitedefaultseppunct}\relax
\EndOfBibitem
\end{mcitethebibliography}

\end{document}